\documentclass[preprint,amssymb,superscriptaddress,longbibliography]{revtex4-1}
\usepackage{graphicx}
\usepackage{amsfonts}
\usepackage{amsmath}
\usepackage[usenames,dvipsnames]{color}
\usepackage{bm}
\usepackage{amssymb}
\usepackage{braket}
\usepackage{dsfont}
\usepackage{comment}

\usepackage{multirow}
\usepackage{epstopdf}
\usepackage[normalem]{ulem}
\usepackage{paralist}

\usepackage{hyperref}
\hypersetup{
colorlinks=true,
linkcolor=blue,
citecolor=blue,
filecolor=green,
urlcolor=blue,
}

\newcommand{\mel}[3]{\langle #1 | #2 | #3 \rangle}

\newcommand{\threejm}[6]{ \left(\begin{array}{ccc} #1 & #3 & #5\\
                                              #2 & #4 & #6
                                \end{array}
                          \right)}
\newcommand{\sixj}[6]{ \left\{\begin{array}{ccc} #1 & #3 & #5\\
                                              #2 & #4 & #6
                                \end{array}
                          \right\}}

\makeatletter
\def\@bibdataout@aps{%
\immediate\write\@bibdataout{%
@CONTROL{%
apsrev41Control%
\longbibliography@sw{%
    ,author="08",editor="1",pages="1",title="0",year="1"%
    }{%
    ,author="08",editor="1",pages="1",title="",year="1"%
    }%
  }%
}%
\if@filesw \immediate \write \@auxout {\string \citation {apsrev41Control}}\fi 
}
\makeatother

\begin{document}
\title{A rigorous adiabatic approach to ultracold atom-molecule collisions in a magnetic field}

\author{Nathan S. Prins}
\affiliation{Department of Physics, University of Nevada, Reno, Nevada, 89557, USA}
\author{Timur V. Tscherbul}
\affiliation{Department of Physics, University of Nevada, Reno, Nevada, 89557, USA}

\date{\today}
\begin{abstract}
    We  extend the rigorous adiabatic coupled-channel formalism  to ultracold nonreactive atom-molecule collisions in the presence of an external magnetic field. The wavefunction of the collision complex is expanded in adiabatic basis states obtained by solving the eigenvalue problem  for the adiabatic  Hamiltonian (the total Hamiltonian of the collision complex minus the  radial kinetic energy) on a grid of  atom-molecule distances $R$. The resulting coupled-channel equations are solved using the diabatic-by-sector method.
    We show that the adiabatic approach provides accurate cross sections for  cold and ultracold Mg ($^1$S) + NH ($^3\Sigma^-$) collisions in a magnetic field with $\simeq$2 times fewer channels than the standard diabatic basis.  We further develop an efficient $R$-dependent basis truncation protocol (RBT), in which  the  elements of  the  log-derivative matrix are sampled and discarded  as  it is propagated from small to large $R$.
While RBT can be applied in both the adiabatic and diabatic bases, we show that the adiabatic basis can be reduced to just the open channels  at long range, leading to an overall computational gain of $\simeq$15{-}30 for the propagation part of the calculation. {The gain is particularly significant in situations where substantial errors in the calculated cross sections ($<$50\%) 
 can be tolerated or long-range interactions are involved, 
 making the adiabatic basis formulation a promising approach to strongly anisotropic collisions and chemical reactions in the presence of an external magnetic field.}
\end{abstract}

\maketitle

\newpage

\section{Introduction}

Ultracold molecular gases are a promising platform for quantum science \cite{Yelin:06,Gorshkov:11,Albert:20,cornish2024}, quantum control of chemical reactions \cite{Krems:08,Balakrishnan:16,Krems:19,Liu:22,karman2024}, and precision tests of fundamental physics \cite{Carr:09,Bohn:17,demille2024}.  
    These applications demand precise quantum control over molecular degrees of freedom, including  their translational motion and internal electronic, rovibrational and hyperfine states. Major experimental advances have been made over the last decade in developing the tools for such control \cite{softley2023,langenYe2024}. In particular, cooling, trapping, and state-selection of molecules  \cite{softley2023,langenYe2024} have enabled  state-to-state tracking of reactants and products in the ultracold KRb~+~KRb $\rightarrow$ K$_2$~+~Rb$_2$ chemical reaction \cite{Liu:21}, observation of a Feshbach resonance in ultracold molecule-molecule collisions \cite{Park:23}, association of ultracold triatomic molecules via atom-molecule Feshbach resonances \cite{Yang:22}, magnetic control of three-body recombination products \cite{Haze:25}, energy transfer from hyperfine to rotational degrees of freedom \cite{Liu:25}, and the realization of a Bose-Einstein condensate of ground-state molecules \cite{Bigagli:24}.

Molecular collisions are central to the ability to control intermolecular interactions in the ultracold regime \cite{Carr:09,Krems:08,Balakrishnan:16,Bohn:17}.
    Ultracold molecular gases typically exhibit universal loss, in which detrimental inelastic collisions occur with near-unit probability once the molecules approach each other at close range \cite{Bause:23}. 
However, elastic collisions are necessary for collisional cooling techniques such as sympathetic and evaporative cooling \cite{Carr:09,Bohn:17,Tscherbul:11,Morita:18,Morita:17}.
Ideally, collisions are expected to be of use as a mechanism to control ultracold molecular gases through Feshbach resonances that are tunable with external fields \cite{Park:23}, as they have been for ultracold atomic gases \cite{Chin:10}. 
Although magnetic Feshbach resonances have been observed in ultracold atom-molecule and molecule-molecule collisions of K~+~NaK \cite{Yang:19,Yang:22,Wang:21}, Na~+~NaLi \cite{Park:23b,Son:22}, and NaLi~+~NaLi \cite{Park:23}, the mechanisms behind these resonances have yet to be fully elucidated because of enormous computational challenges that generally prohibit  rigorous quantum dynamics calculations \cite{Morita:19b,Bause:23,Karman:23}. These challenges stem from the large number of molecular states (rotational, vibrational, fine, and hyperfine) coupled by strongly anisotropic intermolecular interactions \cite{Morita:19b,tscherbulDincao2023}, and the steep scaling of numerically exact coupled-channel calculations (see below), which currently limit the maximum number of  scattering channels to about 18,500 \cite{Suleimanov:12}, with most practical calculations becoming computationally unfeasible already for $>$10,000 channels.
Advancements in molecular quantum scattering theory and computational methods are therefore needed to analyze current and past experiments on ultracold molecular collisions and to provide guidance for future experiments.

The most complete and rigorous theoretical description of molecular collisions is based on the quantum coupled-channel (CC) approach that solves the Schr\"odinger equation numerically exactly for a given Hamiltonian \cite{Child:74,Althorpe:03}. 
    These calculations rigorously account for the coupling of  molecular rotational and orbital angular momenta, as well as  electronic and nuclear spins.
In addition, the inclusion of excited vibrational and electronic states is sometimes necessary \cite{Liu:25}. 
    Highly anisotropic interactions between the collision partners can couple hundreds of rotational states  \cite{moritaTscherbul2024,Liu:25,Karman:23}, necessitating enormous basis sets to achieve convergence.
Because CC calculations scale cubically with the number of basis states (or scattering channels), they tend to be computationally intensive and quickly become intractable as more basis states and/or degrees of freedom are added. 
    
    Fortunately, the efficiency of CC calculations on molecular collisions in external fields can be drastically improved by using optimized channel basis sets, such as those based on the total angular momentum (TAM) \cite{Tscherbul:10,Tscherbul:12} or the total {\it rotational} angular momentum (TRAM) representations \cite{Simoni:06,Chapurin:19,Xie:20,tscherbulDincao2023,moritaTscherbul2024,Liu:25}.
The TRAM is the vector sum of all the angular momenta for mechanical rotation (i.e., the rotation of the diatomic fragment and the orbital motion of the atom in the atom-molecule collision complex). It is rigorously
conserved in the absence of anisotropic spin-dependent interactions, which are often weak compared to the rotational energy splitting and short-range forces \cite{tscherbulDincao2023,moritaTscherbul2024,Liu:25}.  
By leveraging this property, 
 Morita {\it et al.} \cite{moritaTscherbul2024} have  recently succeeded in obtaining converged cross sections for ultracold Rb~+~SrF collisions in a magnetic field based on  {\it ab initio} potential energy surfaces (PESs) in the rigid-rotor approximation. Their CC calculations fully included the hyperfine structure of both Rb and SrF in an external magnetic field, providing the first rigorous insights into magnetic Feshbach resonance spectra in strongly anisotropic atom-molecule collisions \cite{moritaTscherbul2024}.
By contrast, evidence of TRAM non-conservation has been reported in a recent experimental and theoretical study of hyperfine-to-rotational energy transfer in ultracold Rb~+~KRb collisions \cite{Liu:25}. The breakdown of TRAM conservation is likely induced by short-range spin-dependent interactions and conical intersections between the ground and first excited PESs of the Rb-KRb collision complex \cite{Liu:25}.

The CC calculations can also be made more efficient by neglecting certain weak interactions in the short-range region where the anisotropic atom-molecule interactions dominate.
    Recent studies of ultracold Mg~+~NH collisions have found that intramolecular hyperfine \cite{Morita:24} and Zeeman \cite{Morita:24,Vieira:17} interactions can be neglected at short-range, which can be understood using multichannel quantum defect theory with a frame transformation (MQDT-FT) \cite{Morita:24}.
This allows for a several orders-of-magnitude reduction in computational effort compared to full CC calculations including all degrees of freedom \cite{Morita:24}. 
    However, even though MQDT-FT can give quantitatively accurate results for ultracold atom-molecule collisions in a magnetic field without explicitly including the hyperfine structure, its limits  remain to be explored.
For example, the strong short-range coupling between the spin and rotational degrees of freedom in ultracold Rb~+~KRb collisions  \cite{Liu:25} can pose challenges for the MQDT-FT approach, which benefits from neglecting this coupling \cite{Morita:24}.

All previous CC calculations on ultracold atom-molecule and molecule-molecule collisions in magnetic fields have used diabatic basis functions, which are independent of the atom-molecule distance $R$. Examples include the uncoupled space-fixed basis \cite{Volpi:02,Krems:04,Tscherbul:06,GonzalezMartinez:07,Tscherbul:09,Tscherbul:09c,Pavlovic:09,Hummon:11} and  the computationally efficient TAM \cite{tscherbulDalgarno2010,Tscherbul:12,Tscherbul:11,Morita:18,Koyu:22} and  TRAM \cite{tscherbulDincao2023,Liu:25} representations. 
    A disadvantage of the diabatic basis is that it does not account for the peculiar properties of atom-molecule interactions, which typically vary dramatically with $R$. Specifically, at small $R$ these interactions are often strongly anisotropic, coupling a large number of diabatic rotational states \cite{tscherbulDalgarno2010,Tscherbul:12,Tscherbul:11,Morita:18,Morita:24}. At intermediate-to-long range, the anisotropy decreases and the couplings between the basis states decline sharply. A single $R$-independent diabatic basis lacks the flexibility to capture these variations, necessitating the use of extensive diabatic basis sets to achieve convergence of scattering observables, even when the computationally efficient TAM or TRAM representations are used \cite{Tscherbul:11,Morita:18,Morita:24}.

An appealing alternative is offered by the {\it adiabatic basis} composed of the  eigenstates of the full atom-molecule Hamiltonian  evaluated at a fixed value of $R$.
The corresponding eigenvalues, known as the adiabatic potentials, describe how the atom-molecule interaction energy varies as a function of $R$. 
Because the adiabatic basis states explicitly account for the atom-molecule interaction, they are expected to provide a better, more compact description of the quantum states of the collision complex than their diabatic counterparts.  Accordingly, the adiabatic approach is widely used in the theoretical studies of ultracold few-body recombination \cite{greene2017} and reactive scattering \cite{Pack:87,Skouteris:00,Althorpe:03}.   The adiabatic hyperspherical method has enabled rigorous insights into the quantum dynamics of four-body and five-body recombination \cite{DIncao:09,Greene:25} and ultracold chemical reactions Rb~+~Rb~+~Rb $\to$ Rb$_2$~+~Rb \cite{Haze:25} and  K~+~KRb $\to$ K$_2$~+~Rb \cite{daSilva:25}.
However, the adiabatic basis has not yet been used in rigorous calculations of ultracold nonreactive atom-molecule collisions in the presence of an external magnetic field. 

Early implementations of the adiabatic formulation for model atom-molecule collisions in the absence of external fields were based on the R-matrix propagation method \cite{Light:76,Stetchel:1978}.
More recently, log-derivative propagation in the (quasi)-adiabatic basis was implemented in the quantum scattering package MOLSCAT \cite{Hutson:19} even though, to our knowledge,  it has not been used in  practical calculations.
Instead, adiabatic potentials are often used as a valuable tool to analyze the results of CC calculations,  as in exploring the density of magnetic Feshbach resonances  in ultracold Rb~+~SrF collisions \cite{moritaTscherbul2024}, 
     explaining the strong dependence of ultracold reaction rates on the initial states of the reactants  \cite{hermsmeier2021}, and elucidating the mechanisms of rovibrational energy transfer in H$_2$O~+~H$_2$ collisions \cite{Wiesenfeld:21}.   

     A variety of classical and quantum adiabatic capture theories  rely on adiabatic potentials to determine the flux from the initial channel that is transmitted from long-range to short-range \cite{Clary:90,Clary:82,Rackham:03,Alexander:04,Tscherbul:15b,Scribano:18,Pawlak:15,Pawlak:17}.
One can then approximate reaction rates by assuming that the flux is absorbed with unit probability due to the formation of a long-lived complex at short range.
    Statistical adiabatic models are similar to adiabatic capture theories, but assign channel-dependent probabilities for reactive or non-reactive transitions that can take a range of values instead of assuming unit probabilities for all channels \cite{Troe:85,Quack:74,Quack:75}.
     They are also instrumental in analyzing the mechanisms of  microwave and electric field shielding  \cite{Matsuda:20,Anderegg:21,Bigagli:24,Gorshkov:08,Lassabliere:18,Karman:18}.
    Notwithstanding their excellent performance for ultracold collisions of alkali-dimer molecules \cite{Matsuda:20,Anderegg:21,Bigagli:24}, these methods rely on the fundamentally unphysical absorbing boundary condition at short range \cite{Rackham:01} and neglect the non-adiabatic coupling terms,  making  them unsuitable for rigorous quantum dynamics calculations.

In this paper, we develop a rigorous adiabatic methodology for CC calculations on ultracold atom-molecule collisions in a magnetic field. 
We apply the methodology to calculate the cross sections for ultracold Mg~+~NH collisions in a magnetic field, using a realistic {\it ab initio} PES. Significantly, we find that converged scattering cross sections can be obtained
 with far fewer adiabatic channels. 
However, this advantage is somewhat offset by the additional computational effort needed to generate the adiabatic basis states, and to calculate the overlap matrices between the adjacent sectors.  
We also quantify the numerical performance of the adiabatic vs. diabatic basis sets.

In addition, we explore the possibility of truncating the CC basis set ``dynamically'', as the log-derivative matrix is propagated from small to large $R$. 
    {We found that doing so in the adiabatic basis leads to a 15-30 fold reduction in computational cost compared to the diabatic basis for the propagation part of the calculations.}
Our $R$-dependent basis set truncation (RBT) procedure is an  extension of the approach proposed by Light and co-workers  \cite{Stetchel:1978}. The key element of our RBT protocol is that, instead of truncating the intersector overlap  matrix, as done in Ref.~\cite{Stetchel:1978}, we truncate the log-derivative matrix.
    As a result, our method is more general and can be applied in the adiabatic as well as diabatic formulations.

The structure of this paper is as follows. 
    In Sec.~\ref{sec:theory}, we formulate the adiabatic CC theory for ultracold atom-molecule collisions in a magnetic field and 
    describe the RBT  approach.
        In Sec.~\ref{sec:results}, we apply RBT to ultracold Mg~+~NH collisions in a magnetic field and discuss the computational gains achieved over the standard diabatic CC approach.
 Section~\ref{sec:conclusions} concludes by  summarizing the advantages the adiabatic approach has over the diabatic method, and the regimes in which the adiabatic basis is expected to provide the maximal advantage.


\section{Theory} \label{sec:theory}

We begin by presenting the adiabatic coupled-channel (ACC) approach to ultracold atom-molecule collisions in a magnetic field before  describing our $R$-dependent basis truncation (RBT) algorithm. In the next section, the  ACC/RBT approach will be applied to ultracold Mg-NH collisions in a magnetic field. Throughout this paper, we restrict attention to nonreactive atom-molecule collisions in a magnetic field. The reader is referred to Refs.~\cite{Tscherbul:15b,Tscherbul:25} for an extension of the adiabatic approach to chemically reactive scattering in an electric field using hyperspherical  coordinates \cite{Skouteris:00}.

\subsection{Hamiltonian and adiabatic coupled-channel (ACC) equations}

The Hamiltonian for an atom-molecule collision complex A~+~BC can be written as (in atomic units) 
\cite{Child:74,Volpi:02,Krems:04}
\begin{equation}
    \hat{H} ({\bf R},{\bf r}) = -\frac{1}{2\mu R} \frac{\partial^2}{\partial R^2} R + \hat{H}_{\text{ad}} ({\bf R},{\bf r}),
    \label{eq:fullHamiltonian}
\end{equation}
where $\mu$ is the reduced mass of the complex, and ${\bf R}$ and ${\bf r}$ are the standard Jacobi vectors:  
  ${\bf R}$ extends from the center of mass of BC to A  and ${\bf r}$ joins the nuclei of  BC. 
  The adiabatic Hamiltonian 
        \begin{equation}
            \hat{H}_{\text{ad}} ({\bf R}, {\bf r}) = \frac{\hat{L}^2}{2 \mu R^2} + \hat{V}({R}, {r},\theta) + \hat{H}_{\text{as}}({\bf r}),
            \label{eq:adiabaticHamiltonian}
        \end{equation}
 includes, in order, the centrifugal kinetic energy due to the orbital motion of A around BC, the interaction potential describing the interaction between the atom and molecule, and the asymptotic Hamiltonian that describes the isolated (i.e., non-interacting) atom and molecule, defined as
\begin{equation}
    \hat{H}_{\text{as}}({\bf r}) = \lim_{R \rightarrow \infty} \hat{H}_{\text{ad}} ({\bf R},{\bf r}).
    \label{eq:asymptoticHamiltonianLimit}
\end{equation}
In Eq.~\eqref{eq:adiabaticHamiltonian}, $\hat{L}$ is the orbital angular momentum that describes rotations of $\bf{R}$,
$R=|\mathbf{R}|$, $r=|\mathbf{r}|$, and $\theta$ is the angle between the vectors $\mathbf{R}$ and $\mathbf{r}$. 
We assume that the A-BC interaction is adequately described by a single adiabatic potential energy surface (PES), and freeze the internuclear distance of the diatomic molecule at its equilibrium value, $r=r_e$. The rigid-rotor approximation is known to be reliable for rigid diatomic molecules such as NH colliding in their ground vibrational states \cite{Volpi:02,Krems:04}.

Here, we consider the prototypical example of a binary collision between a $^3\Sigma$ diatomic molecules and a structureless $^1$S-state atom in an  external magnetic field ${\bf B}$. 
The asymptotic Hamiltonian then reduces to that of an isolated  $^3\Sigma$ molecule such as the NH radical \cite{Krems:04} 
        \begin{equation}
            \hat{H}_{\text{as}} ({\bf r}) = B_{\text{e}} \hat{{\bf N}}^2 + \gamma_{\text{sr}} \hat{{\bf N}} \cdot \hat{{\bf S}} + 2 \mu_0 {\bf B} \cdot \hat{{\bf S}} + \hat{V}_{\text{SS}}
            \label{eq:asymptotic_hamiltonian}
        \end{equation}
where $\hat{{\bf N}}$ and $\hat{{\bf S}}$ are the angular momentum operators that describe molecular rotation and electronic spin, respectively, $B_e$ is the rotational constant, $\gamma_{\text{sr}}$ the the spin-rotation constant, $\mu_0$ is the Bohr magneton, and $\hat{V}_{\text{SS}}$ is the spin-spin interaction \cite{Krems:04}.

Following the general strategy of the adiabatic approach \cite{Child:74,greene2017,DIncao:09, Wang:11},  we expand the wavefunction of the triatomic atom-molecule collision complex at a total energy $E$ in the adiabatic basis as
\begin{equation}
    \psi_{i_0} ({\bf R},{\bf r}) = \frac{1}{R} \sum_{i=1}^{\mathcal{N}} F_{ii_0}(R) \Phi_i({\bf r},\hat{R};R).
    \label{eq:adiabaticexpansion}
\end{equation}
Here, $F_{ii_0}(R)$ are the radial solutions of the time-independent Schr\"odinger equation $\hat{H}\psi_{i_0}=E\psi_{i_0}$ which form a square $\mathcal{N}\times \mathcal{N}$ matrix ${\bf F}(R)$, whose 
columns represents linearly independent radial solutions  for a given initial state $i_0$ \cite{Child:74}.

The adiabatic basis functions $\Phi_i$ are eigenstates of the adiabatic Hamiltonian
\begin{equation}
    \hat{H}_{\text{ad}} \Phi_i({\bf r},\hat{R};R) = \epsilon_i(R) \Phi_i({\bf r},\hat{R};R),
    \label{eq:adiabaticEigenvalueProblem}
\end{equation}
where $\epsilon_i(R)$ and $ \Phi_i({\bf r},\hat{R};R)$ are the eigenvalues of the adiabatic Hamiltonian, also known as adiabatic potentials, and $\Phi_i({\bf r},\hat{R};R)$ are the corresponding eigenfunctions, also known as adiabatic surface functions \cite{Pack:87,Skouteris:00}. 
The adiabatic eigenvalues and eigenfunctions depend on $R$ only parametrically,  and  approach those of the asymptotic Hamiltonian as $R \rightarrow \infty$, which is a direct consequence of Eq.~\eqref{eq:asymptoticHamiltonianLimit}.
The adiabatic eigenvalue problem \eqref{eq:adiabaticEigenvalueProblem} is solved at a fixed value of the atom-molecule distance $R$ by expanding the adiabatic eigenfunctions in a primitive $R$-independent basis set
\begin{equation}
    \Phi_{i} ({\bf r},\hat{R}; R) = \sum_{j=1}^{\mathcal{N}} T_{j i}(R) \chi_j({\bf r},\hat{R}) ,
    \label{eq:adiabatic_eigenstates}
\end{equation}
where $\mathcal{N}$ is the number of primitive basis functions used to in the expansion.
The expansion coefficients $ T_{j i}(R)$ are found by solving the  matrix eigenvalue problem 
\begin{equation}
    \sum_{j=1}^{\mathcal{N}} \left( \mel{\chi_i}{\hat{H}_{\text{ad}} }{\chi_j} - \epsilon_j(R) \delta_{ij} \right) T_{j i} (R) = 0
    \label{eq:matrixEigenvalueProblem}
\end{equation}
obtained by substituting Eq.~\eqref{eq:adiabatic_eigenstates}  into Eq.~\eqref{eq:adiabaticEigenvalueProblem}.
The expansion coefficients $ T_{j i}(R)$ form columns of the eigenvector  matrix ${\bf T}(R)$, which diagonalizes the matrix of the adiabatic Hamiltonian in the primitive basis with elements $\mel{\chi_i}{\hat{H}_{\text{ad}} }{\chi_j}$.

To solve the adiabatic eigenvalue problem for our triatomic   complex composed of a $^3\Sigma$ molecule colliding with a $^1$S atom, we use
the uncoupled space-fixed (SF) representation \cite{Volpi:02,Krems:04} as our primitive basis
\begin{equation}
    \ket{\chi_j} = \ket{N M_N} \ket{S M_S} \ket{L M_L}.
    \label{eq:uncoupledSF}
\end{equation}
These basis states are simultaneous eigenstates of $\hat{N}^2$, $\hat{N}_z$, $\hat{S}^2$, $\hat{S}_z$, $\hat{L}^2$, and $\hat{L}_z$.
The SF quantization axis points along the magnetic field vector such that ${\bf B} = B \hat{z}$. The uncoupled SF basis is well suited for weakly and moderately anisotropic interaction potentials. For strongly anisotropic interactions, the total angular momentum \cite{Tscherbul:10} or total rotational angular momentum \cite{tscherbulDincao2023} bases are much more  computationally efficient.

The matrix elements of the adiabatic Hamiltonian  in the primitive basis $\mel{\chi_i}{\hat{H}_{\text{ad}} }{\chi_j}$ required to set up the matrix eigenvalue problem \eqref{eq:matrixEigenvalueProblem} are obtained as described in Ref.~\cite{Krems:04}. These include the matrix elements of the asymptotic Hamiltonian, the centrifugal kinetic energy, and the atom-molecule interaction potential from Eq.~\eqref{eq:adiabaticHamiltonian}
\begin{equation}
\mel{\chi_i}{\hat{H}_{\text{ad}} }{\chi_j}
=\bra{N M_N} \bra{S M_S} \bra{L M_L}  
\biggl{(}
\frac{\hat{L}^2}{2 \mu R^2} + \hat{V}({R},\theta) + \hat{H}_{\text{as}} \biggr{)}
            \ket{N' M_N'} \ket{S M_S'} \ket{L' M_L'}
\end{equation}
All of these matrix elements can be evaluated as described in the seminal papers by Volpi and Bohn \cite{Volpi:02} and Krems and Dalgarno~\cite{Krems:04}.
For example, to compute the matrix elements of the atom-molecule interaction potential, we expand it in Legendre polynomials 
        \begin{equation}
            V(R,\theta) = \sum_{\lambda} V_{\lambda}(R) P_{\lambda}( \cos{\theta} ),
            \label{eq:legendre_expansion}
        \end{equation}
where $V_{\lambda}$ are the $R$-dependent expansion coefficients, and  use Eq.~(12) of Ref.~\cite{Krems:04}.

As a final step, to obtain the adiabatic radial solutions $F_{ii_0}(R)$ in Eq.~\eqref{eq:adiabaticexpansion} we plug Eq.~\eqref{eq:adiabaticEigenvalueProblem} into the time-independent Schr\"odinger equation, which  
yields the coupled-channel equations in the adiabatic basis (the ACC equations) \cite{Child:74,Wang:11,greene2017,DIncao:09} 
        \begin{align}\notag
            \biggl{[} \frac{d^2}{dR^2} + 2\mu (E - \epsilon_i(R)) \biggr{]} F_{ii_0}(R) = \sum_j \biggl{[} &-\langle \Phi_i({\bf r},\hat{R};R) \biggl{|}\frac{\partial}{\partial R} \biggl{|}\Phi_j({\bf r},\hat{R};R) \rangle
            \frac{\partial}{\partial R} 
            \\ & - \langle\Phi_i({\bf r},\hat{R};R) \biggl{|}\frac{\partial^2}{\partial R^2} \biggl{|}\Phi_j({\bf r},\hat{R};R)\rangle \biggr{]} F_{ji_0}(R),
            \label{eq:ccWithCoupling}
        \end{align}
        where integration in the radial derivative  coupling matrix elements $\langle ...\rangle$ is performed over $\mathbf{r}$.
The boundary conditions on the radial solutions ${\bf F}(R)$ are 
\begin{equation}
   \begin{aligned}
      \lim_{R \to 0} F_{ii_0}(R) =     & \hspace{1.5mm} 0 \\
      \lim_{R\to\infty} F_{ii_0}(R) =  & \hspace{1.5mm} \delta_{ii_0} \exp[-i(k_{i_0} R - \pi l_0/2)] \\
                                    & - \left( \frac{k_{i_0}}{k_i} \right)^{1/2} S_{ii_0} \exp[-i(k_i R - \pi l_i/2)],
      \label{eq:boundary_conditions}
   \end{aligned}
\end{equation}
where $S_{ii_0}$ is the scattering matrix element and $l_i$ and $k_i=\sqrt{2\mu (E - \epsilon_i)}$ are the orbital angular momentum quantum number and wavevector of the $i^{\text{th}}$ adiabatic channel, respectively ($l_0$ and $k_{i_0}$ for the initial channel). In the limit $R\to\infty$ the adiabatic potentials tend to the eigenvalues of $\hat{H}_\text{as}$, which define the internal  (rovibrational and Zeeman) energy levels of the isolated diatomic molecule $\epsilon_i=\epsilon_i(R\to\infty)$.
We will assume that the initial state is fixed throughout, and omit the subscript $i_0$ from now on.

\subsection{The diabatic-by-sector formulation}
        
The ACC equations~\eqref{eq:ccWithCoupling} include the matrix elements of the radial derivative operators $\frac{\partial}{\partial R}$ and $\frac{\partial^2}{\partial R^2}$, which couple different adiabatic states. These non-adiabatic couplings  
are challenging to handle numerically because they diverge at avoided crossings \cite{Child:74}, which are common in atom-molecule systems (see below).

To avoid dealing with the non-adiabatic couplings,  
  we use the diabatic-by-sector (DBS) approach  
commonly employed in quantum reactive scattering calculations 
\cite{Pack:87,Skouteris:00},
in which the entire range of $R$ is split into small sectors. Within each sector, the adiabatic states are independent of $R$, so non-adiabatic couplings can be neglected 
and Eq.~\eqref{eq:ccWithCoupling} becomes
        \begin{equation}
            \left[ \frac{d^2}{dR^2} + 2\mu E - 2\mu \epsilon_i(R) \right] F_{ii_0}(R) = 0.
            \label{eq:ccDBS}
        \end{equation} 
The log-derivative matrix is propagated   within each sector using Eq.~\eqref{eq:ccDBS} and transformed from the adiabatic basis of the current sector to that of the next by a unitary transformation. At the final atom-molecule separation ($R = R_{\text{final}}$), the log-derivative is transformed to the eigenbasis of the asymptotic Hamiltonian $\hat{H}_{\text{as}}$ and scattering boundary conditions are applied to obtain the $S$-matrix. 
The details of this procedure are described below. 

\begin{figure}[t!]
    \centering
    \includegraphics[width=1.0\textwidth]{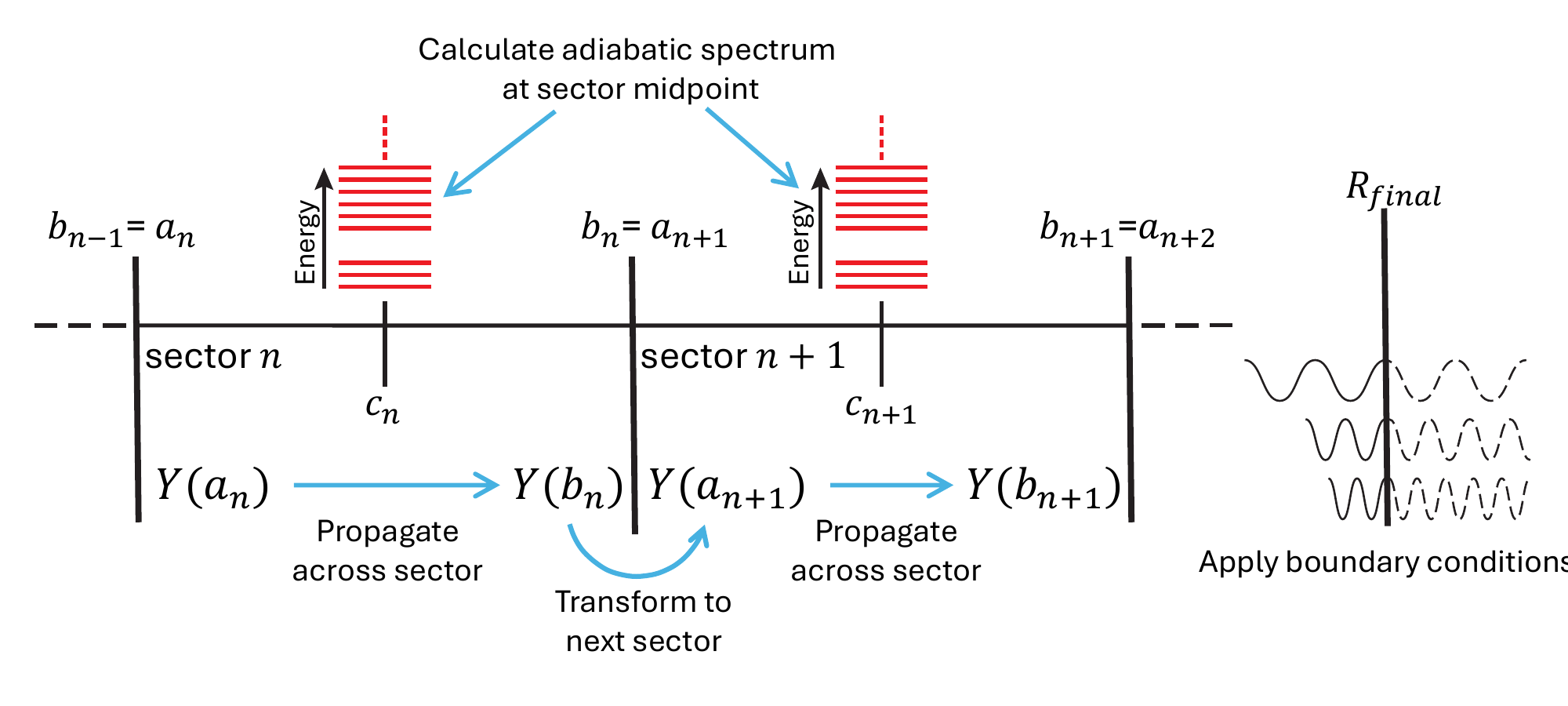}
    \caption{Schematic of the diabatic-by-sector procedure for  propagating the log-derivative matrix in the adiabatic basis.  The $n^{\text{th}}$ sector is defined on an interval $R \in [a_n,b_n]$ by the starting point $R=a_n$, the midpoint $R=c_n$, and the endpoint $R=b_n$. The endpoint of the $n^{\text{th}}$ sector is the starting point of the following sector ($a_n = b_{n-1}$). The blue arrows show the propagation and sector-to-sector transformation of the log-derivative matrix.}
    \label{fig:sector_diagram}
\end{figure}

As shown in Fig.~\ref{fig:sector_diagram}, the $n^{\text{th}}$ sector is defined by the initial point $R=a_n$, the midpoint $R=c_n$, and the endpoint $R=b_n$.
The adiabatic eigenvalue problem is solved at the midpoint of each sector  by expanding the adiabatic states in the primitive diabatic basis [Eqs. \eqref{eq:adiabatic_eigenstates} and \eqref{eq:uncoupledSF}]. The matrix elements of the adiabatic Hamiltonian in this basis are evaluated analytically as described in, e.g., Ref.~\cite{Krems:04}.  Numerical diagonalization of the resulting matrix provides the adiabatic energies and eigenvectors as a function of $R$. The former are used in  Eq.~\eqref{eq:ccDBS} whereas the latter are employed to construct the sector-to-sector overlap matrix as described below.

The log-derivative matrix in the adiabatic basis of the $n^{\text{th}}$ sector defined at an $R$-value within the sector ($R=R_n$) is 
\begin{equation}
    {\bf Y}(R_n) = {\bf F}'(R_n) [{\bf F}(R_n)]^{-1},
    \label{eq:log_derivative}
\end{equation}
The subscript $n$ denotes the adiabatic basis in which the log-derivative is represented.
To propagate the log-derivative matrix within a sector from $R=a_n$ to $R=b_n$, we use the methods proposed by Johnson and Manolopoulos \cite{Johnson:73,Manolopoulos:86}.
Following Ref.~\cite{Manolopoulos:86}, we define the reference potential of the $n^{\text{th}}$ sector in the adiabatic basis defined at $R=c_n$  as
\begin{equation}
    {\bf W}_{\text{ref}} (c_n) = 2\mu {\bf T}_n^{\text{T}} {\bf H}_{\text{ad}}(c_n) {\bf T}_n - 2\mu E {\bf 1},
    \label{eq:reference_potential}
\end{equation}
where ${\bf 1}$ is the identity matrix, ${\bf T}_n$ is the adiabatic eigenvector matrix of sector $n$, and ${\bf T}_n^{\text{T}} {\bf H}_{\text{ad}} (c_n){\bf T}_n$ is the diagonal matrix of adiabatic eigenvalues $\epsilon_i(c_n)$. 

Note that ${\bf H}_{\text{ad}}$ is the adiabatic Hamiltonian written in the primitive basis with matrix elements $[ {\bf H}_{\text{ad}} ]_{ij} = \mel{\chi_i}{\hat{H}_{\text{ad}}}{\chi_j}$.
The residual coupling matrices at any value of $R$ in the $n$-th sector are
\begin{equation}
    {\bf U} (R) =  2\mu {\bf T}_n^{\text{T}} \left[ {\bf H}_{\text{ad}} (R) - {\bf H}_{\text{ad}} (c_n) \right] {\bf T}_n.
    \label{eq:residual_coupling_matrix}
\end{equation}
In the calculations presented throughout this paper, we avoid this transformation by reducing the width of the sectors small enough to validate the approximation that the adiabatic basis is constant throughout the sector.

The residual coupling matrix vanishes at the sector midpoint ($R_n=c_n$), so the quadrature matrix in the improved log-derivative method \cite{Manolopoulos:86} 
\begin{equation}
    {\bf Q}(c_n) = \frac{4}{h} \left[ {\bf 1} - \frac{h^2}{6} {\bf U}(c_n) \right]^{-1} - \frac{4}{h} {\bf 1},
    \label{eq:qc}
\end{equation}
also vanishes, eliminating one inversion that is required when using a diabatic basis. 
In Eq.~\eqref{eq:qc}, $h$ is the half-width of the sector. 
The trade-off for eliminating the inversion 
is the diagonalization that needs to be performed to solve the adiabatic eigenvalue problem (Eq.~\eqref{eq:adiabaticEigenvalueProblem}).
We note that 
this inversion, like the diagonalization required for the adiabatic treatment, is independent of the total energy, and therefore only needs to be calculated when the adiabatic Hamiltonian changes, like when changing the magnetic field strength or scaling the interaction potential. 

Once the log-derivative matrix is propagated across the sector, ${\bf Y}(a_n) \rightarrow {\bf Y}(b_n)$, it is transformed to the adiabatic basis of the next sector as \cite{Pack:87}
\begin{equation}
    {\bf Y}(a_{n+1}) = {\bf O}_{n \rightarrow n+1}^{\text{T}} {\bf Y}(b_n) {\bf O}_{n \rightarrow n+1},
    \label{eq:ld_transformation}
\end{equation}
where the overlap matrix is computed from the adiabatic eigenvector matrices from subsequent sectors:
\begin{equation}
    {\bf O}_{n \rightarrow n+1} = {\bf T}_n^{\text{T}} {\bf T}_{n+1}.
    \label{eq:overlap}
\end{equation} 
The sector-to-sector similarity transformation  \eqref{eq:ld_transformation}
can be regarded as consisting of two steps, the first of which transforms the log-derivative matrix from the adiabatic basis of the $n^\text{th}$ sector to the primitive basis, and the second from the primitive basis to the adiabatic basis of the $(n+1)^\text{th}$ sector. 

On reaching the last propagation sector centered at $R_n=R_{N_\text{s}}$ in the asymptotic region of large $R$,  
the log-derivative matrix is transformed back to the diabatic basis, and then to a basis, which diagonalizes the asymptotic Hamiltonian \eqref{eq:asymptoticHamiltonianLimit} \cite{Krems:04}.
The asymptotic transformation matrix is a product of two orthogonal matrices 
\begin{equation}
    {\bf \mathbf{O}}_{N_{\text{s}} \rightarrow \text{as}} = {\bf T}_{N_{\text{s}}}^{{T}} {\bf C},
\end{equation}
where $\mathbf{C}$ is the matrix of eigenvectors of the asymptotic Hamiltonian.
As this transformation brings the log-derivative matrix to the proper diabatic basis of eigenstates of the asymptotic Hamiltonian, one can now apply the standard K-matrix  boundary conditions as described by Krems and Dalgarno \cite{Krems:04} to obtain  
the K and S-matrices using the standard expressions, e.g., \cite{Johnson:73}. 
    The key observables of interest --- the integral cross sections and transition probabilities  between the different internal molecular levels induced by atom-molecule collisions --- are obtained from the S-matrix using well-documented expressions \cite{Child:74,Krems:04}.

\subsection{Truncating the adiabatic basis during propagation} \label{sec:RBT}

As noted above, rigorous CC calculations of non-reactive molecular  collisions in external fields typically use the same $R$-independent (diabatic) basis throughout the entire range of $R$.
A more computationally efficient procedure would be to reduce the size of the basis (diabatic or adiabatic) during propagation.   Stetchel, Walker, and Light \cite{Stetchel:1978}  implemented  such  a procedure by sampling the elements of the sector-to-sector overlap matrix in the adiabatic basis \eqref{eq:overlap} but provided no details on either the implementation or performance of the method. In addition, their approach cannot be implemented in the diabatic basis, for which the sector-to-sector overlap matrix reduces to the identity matrix.
    In this section, we propose and describe an  alternative  method based on sampling the log-derivative matrix, which works in {\it both the diabatic and adiabatic formulations}.

In the remainder of this paper, we will refer to the general class of methods that reduce the size of the basis during propagation as $R$-dependent basis truncation (RBT). 
    We refer to our version as log-derivative-based RBT and the version presented by Stetchel, Walker, and Light as overlap-based RBT.
Below, we focus on log-derivative-based RBT in the adiabatic basis applied to ultracold atom-molecule collisions.
    Additional details regarding the performance of RBT
    are relegated to Appendix~\ref{app:RBT}.
    The RBT procedure in the diabatic basis, along with the original overlap-based RBT \cite{Stetchel:1978}, is described in Appendix~\ref{app:diabatic_RBT}.

As noted above, in log-derivative-based RBT we sample the elements of the log-derivative matrix  as it is propagated from small to large $R$.
    In each propagation sector centered at $R=c_n$, the propagation basis size $M_n$ defines the dimension of the log-derivative matrix ($\text{dim}({\bf Y}) = M_n \times M_n$).
Note that the propagation basis size is different from the total basis size $\mathcal{N}$ used to solve the adiabatic eigenvalue problem defined by  Eq.~\eqref{eq:adiabaticEigenvalueProblem}.
    Our aim is to reduce the propagation basis size to lower the computational cost of matrix operations
required to propagate the log-derivative matrix across a sector  and to transform it to the next sector.

In a given sector, the $i^{\text{th}}$ channel is locally open if $\epsilon_i(R)<E$ and locally closed otherwise, where $E$ is the total energy.
Let $M^{(\text{LO})}_{n}$ and $M^{\text{(LC)}}_{n}$ be the numbers of locally open and locally closed channels in the $n^{\text{th}}$ sector, respectively.
A weakly closed channel is any closed channel that becomes locally open in any sector, i.e., satisfies $\epsilon_i(c_n) < E$ for at least one value of $c_n$.
We start RBT at the $R$-value corresponding to  the minimum of the isotropic part of the interaction potential, $V_0(R)$.
After that condition is met, the following algorithm describes how the log-derivative matrix is sampled to determine how many channels can be removed in the $n^{\text{th}}$ sector. 
    The threshold $input\_threshold$ is a convergence  parameter.

    The outermost {\bf for $i$} loop  runs over the  locally closed channels, starting with the channel with the greatest adiabatic energy in the $n^{\text{th}}$ sector (we assume that the adiabatic eigenvalues are sorted in the order of increasing energy). 
The next {\bf for $j$} loop runs over the locally open channels and computes the maximum absolute value ($max_{\_}value$) of the log-derivative matrix element between the $i^{\text{th}}$ closed channel and the locally open channels.

The following {\bf if ... then} statement checks whether $max\_value$
is less than or equal to a predetermined threshold $input\_threshold$. If so, the closed channel is removed, and the propagation basis size of the following sector $M_{n+1}$ is reduced by one.
Otherwise, the closed channel is kept, and the log-derivative is transformed to the next sector as usual. 

Once the RBT algorithm determines which channels are to be removed, the corresponding rows and columns of the log-derivative matrix are deleted in the sector-to-sector transformation step, in which the log-derivative matrix is transformed from the basis of one sector to that of the following sector [Eq.~\eqref{eq:ld_transformation}].


\vspace{4mm}
\noindent \hrule
\vspace{1.5mm}
\noindent \hrule
\vspace{2mm}

\noindent $M_{n+1} = M_n$

\noindent {\bf for} $i$ from $M_n$ to $M_n - M^{\text{(LC)}}_{n} + 1$

\noindent \hspace{5mm} $max\_value$ = 0

\noindent \hspace{5mm} {\bf for} $j$ from 1 to $M^{(\text{LO})}_{n}$

\noindent \hspace{10mm} $matrix\_element = \left[ {\bf Y}(b_n) \right]_{ji}$

\noindent \hspace{10mm} Find the maximal $|matrix\_element|$ $\to$ $max\_value$

\noindent \hspace{5mm} {\bf end if}

\noindent \hspace{5mm} {\bf if} $max\_value \leq input\_threshold$ {\bf then}

\noindent \hspace{10mm} remove the $i^{\text{th}}$ channel

\noindent \hspace{10mm} $M_{n+1} = M_{n+1} - 1$

\noindent \hspace{5mm} {\bf else}

\noindent \hspace{10mm} exit loop over $i$

\noindent \hspace{5mm} {\bf end if}

\noindent {\bf end for}

\vspace{3mm}
\noindent \hrule
\vspace{0.5mm}
\noindent \hrule

\begin{figure}[t!]
    \centering
    \includegraphics[width=0.45\textwidth]{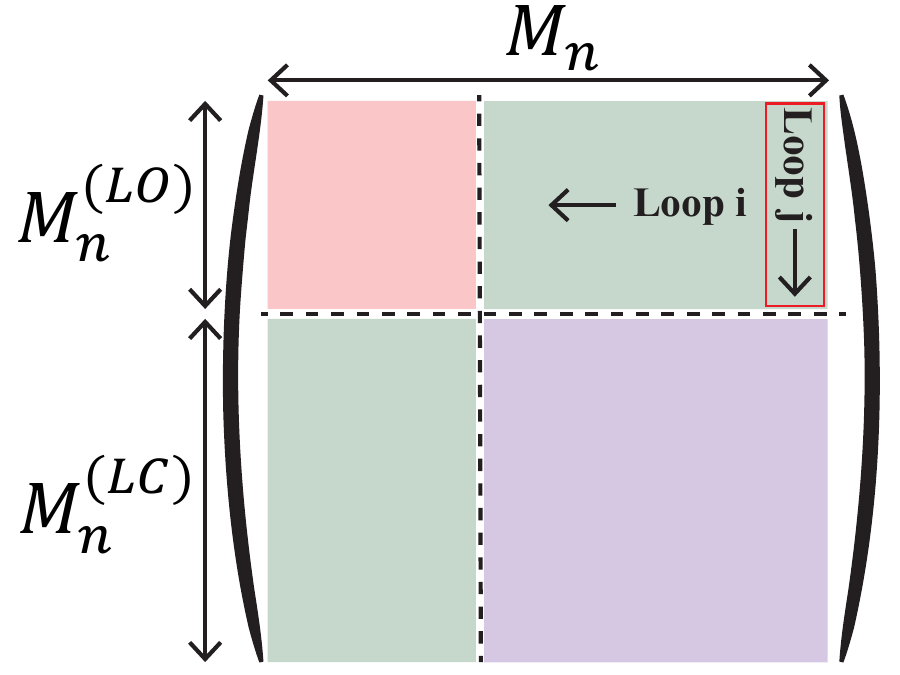}
    \caption{Schematic representation of the log-derivative matrix (${\bf Y}$) in the adiabatic basis as a square matrix of dimension $M_n \times M_n$. The log-derivative matrix is divided into four submatrices whose dimensions are based on the number of locally open channels ($M^{\text{(LO)}}_n$) and the number of locally closed channels ($M^{\text{(LC)}}_n$). The matrix is symmetric, so the green sections are transposes of one another.}
    \label{fig:LD_diagram}
\end{figure}

For example, suppose we find that the basis can be reduced from $M_n$ to $M_{n+1}<M_n$ after propagating the log-derivative through the $n^{\text{th}}$ sector.
The $M_n \times M_{n+1}$ overlap matrix ${\bf \mathbf{O}}_{n \rightarrow n+1} = {\bf T}_{n}^{{T}} {\bf T}_{n+1}$ is then constructed from the first $M_n$ columns of ${\bf T}_n$ and the first $M_{n+1}$ columns of ${\bf T}_{n+1}$.
Using this overlap matrix to transform the log-derivative matrix via Eq.~\eqref{eq:ld_transformation} reduces the dimension of the log-derivative matrix from $M_n \times M_n$ to $M_{n+1} \times M_{n+1}$.

The above truncation procedure progressively reduces the size of the log-derivative matrix as it is propagated from small to large $R$. This, in turn, reduces the cost of the two inversions required to propagate the log-derivative matrix across a sector from $O(\mathcal{N}^3)$ to $O(M_n^3)$. 


\section{Results} \label{sec:results}

\subsection{Adiabatic coupled-channel calculations for ultracold Mg~+~NH collisions in a magnetic field}

In this section, we apply the rigorous adiabatic approach developed in Sec.~II to calculate the cross sections for ultracold Mg~+~NH collisions in a magnetic field. 
The Mg-NH interaction is moderately anisotropic, enabling fully converged CC calculations using standard diabatic basis sets \cite{Vieira:17,Morita:24,Wallis:09,GonzalezMartinez:11}.
This makes ultracold Mg~+~NH collisions a convenient test system, allowing us to benchmark the adiabatic approach against the standard diabatic CC calculations.

\begin{figure}[t!]
    \centering
    \includegraphics[width=0.45\textwidth]{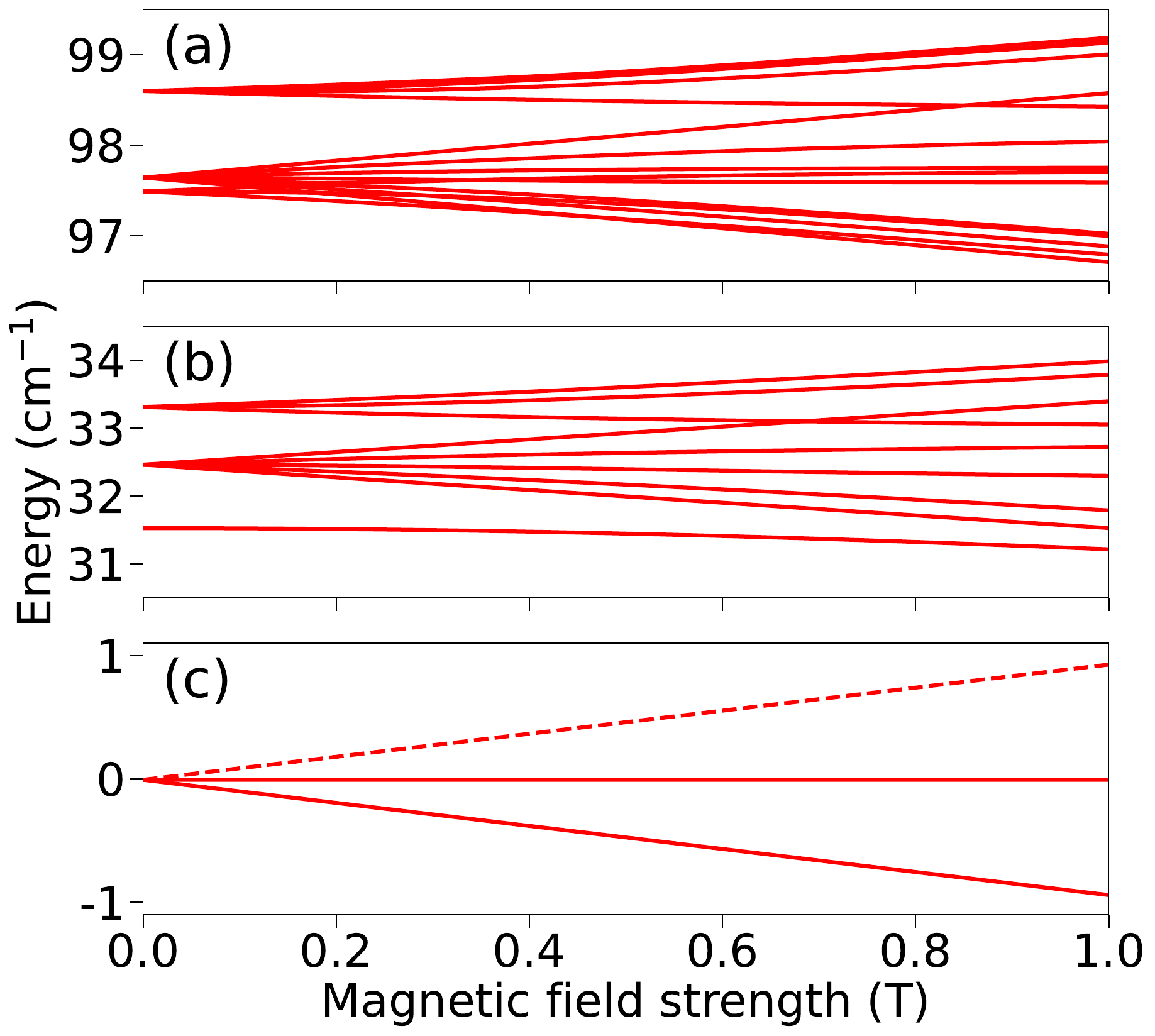}
    \caption{The $N=0$ (a), $N=1$ (b), and $N=2$ (c) energy levels of NH ($^3 \Sigma ^-$) plotted against the magnitude of the applied magnetic field. The initial state for scattering calculations is shown by the dashed line.}
    \label{fig:isolated_spectrum}
\end{figure}

\begin{figure}[h!]
    \centering
    \includegraphics[width=0.45\textwidth]{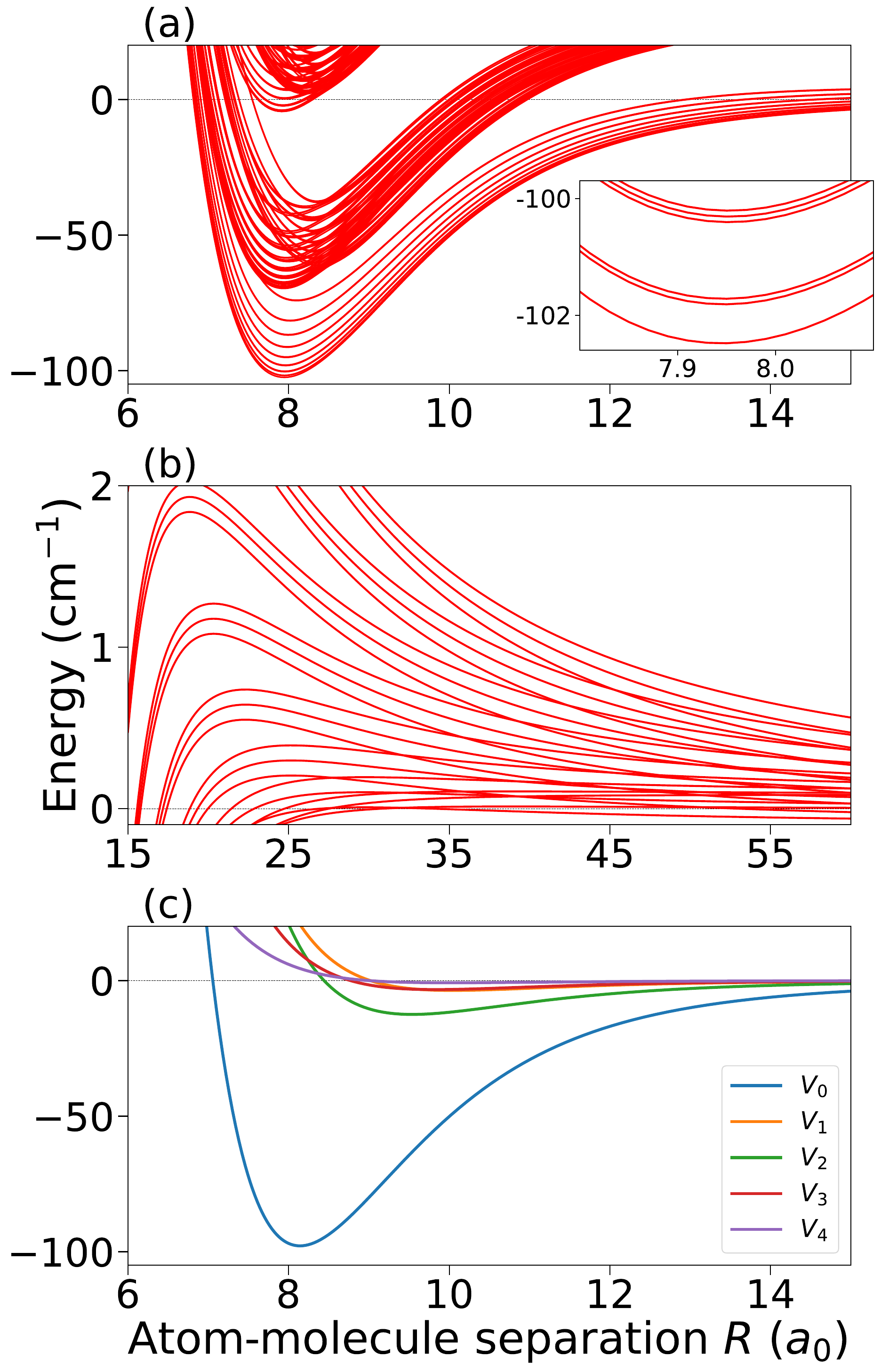}
    \caption{(a) Lowest adiabatic potentials of the Mg-NH collision complex. The inset zooms into the minimum of the six deepest adiabatic potentials to highlight the Zeeman splitting. (b) Medium- to long-range adiabatic potentials of the Mg-NH collision complex. All potentials have a well-defined value of $M_{\text{tot}} = 1$ and are calculated at a magnetic field of $B = 1000$~G. The zero of energy corresponds to the asymptotic value of the Mg-NH interaction potential ($R \rightarrow \infty$). (c) Legendre expansion coefficients of the Mg-NH interaction potential. }
    \label{fig:potential_mg-nh}
\end{figure}

Figure~\ref{fig:isolated_spectrum} shows the magnetic field dependence of the lowest rotational-Zeeman levels of an isolated NH($^3\Sigma$) molecule.
The rotational levels are arranged in manifolds, with the rotational spacing between the manifolds being much larger than the intramolecular spin-spin interaction \cite{Krems:04,Brown:03}.
    We can therefore label each manifold by its  rotational quantum number $N$. 
The dashed line shows the low-field seeking Zeeman energy level of the ground rotational state, which is the initial state used in all calculations presented here.
    Inelastic transitions from this state (also known as Zeeman or spin relaxation) cause loss of NH molecules from a magnetic trap \cite{Krems:04,Krems:03b,Campbell:07,Campbell:09}.

Figures~\ref{fig:potential_mg-nh}(a,b) show the adiabatic potentials for the Mg~+~NH complex. 
    Because the Mg-NH interaction is dominated by the isotropic  $V_0(R)$ term in Eq.~\eqref{eq:legendre_expansion}, as shown in Fig.~\ref{fig:potential_mg-nh}(c), the deepest adiabatic potentials closely follow the shape of $V_0(R)$.
The avoided crossings at short-range are mediated by the anisotropic part of the interaction potential. 
    At long-range, the avoided crossings occur due to the different values of $L$ in each adiabatic channel, resulting in long-range centrifugal barriers being shifted with respect to each other.
Significantly, the presence of avoided crossings at both short- and long-range suggests that the non-adiabatic coupling terms in Eq.~\eqref{eq:ccWithCoupling} cannot be neglected at any value of $R$.
    The non-adiabatic couplings must be accounted for using, e.g., the diabatic-by-sector (DBS) method as described below.

\begin{figure}[t!]
    \centering
    \includegraphics[width=0.45\textwidth]{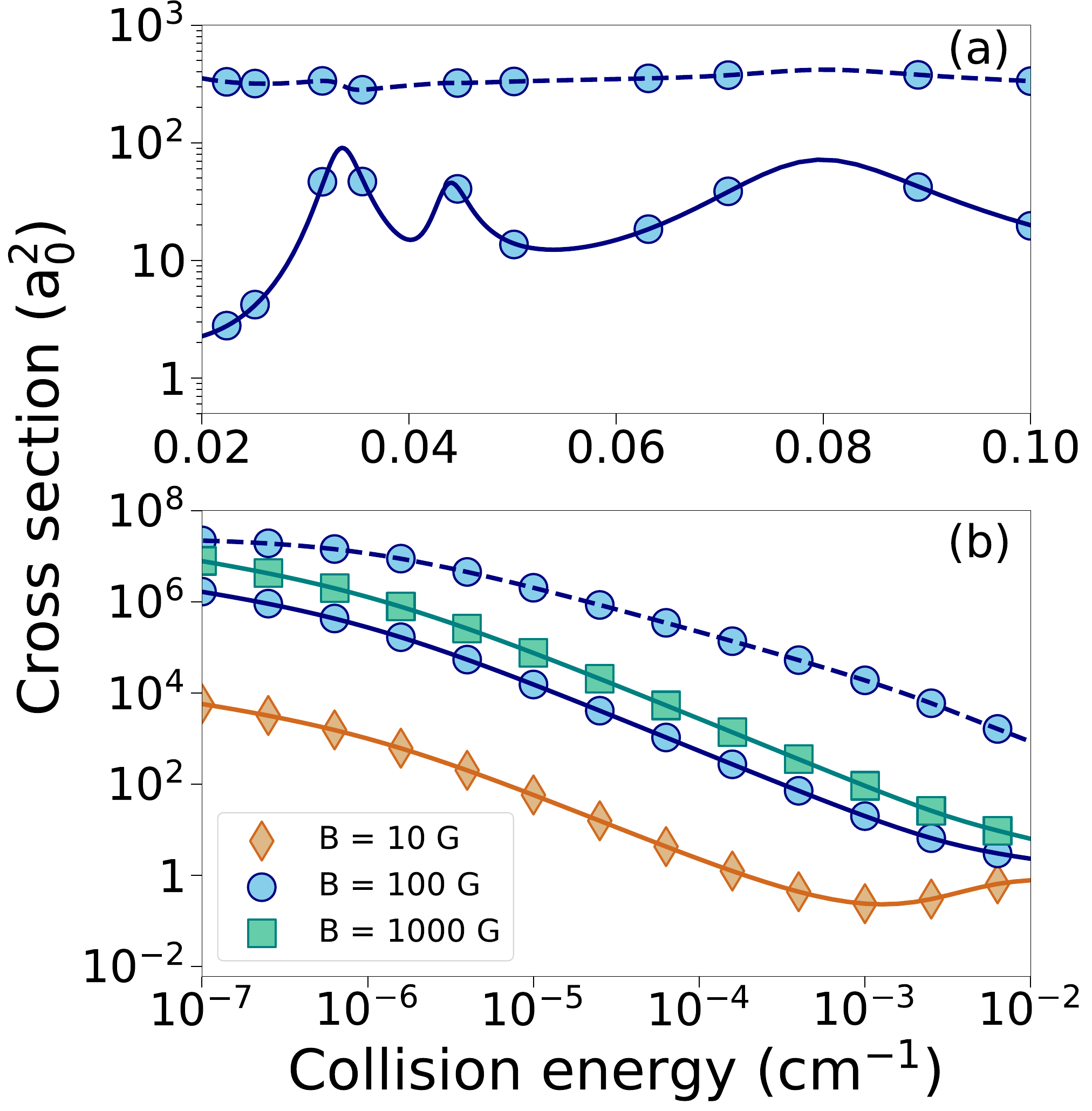}
    \caption{Elastic (dashed lines) and total inelastic (solid lines) cross sections as a function of collision energy computed for ultracold Mg~+~NH collisions calculated via the adiabatic formulation using $R$-dependent truncation with a threshold of $0.001 \text{ }a_0^{-1}$. Open shapes are cross sections calculated using the standard diabatic treatment with the uncoupled SF representation at magnetic field strengths 10~G (brown diamonds), 100~G (navy circles), 1000~G (teal squares). The upper panel (a) shows cross sections near a resonance at $B=100$~G, and the lower panel (b) shows cross sections at ultracold energies. }
    \label{fig:autotrunc_accuracy}
\end{figure}

Figure~\ref{fig:autotrunc_accuracy} shows the results of our CC calculations for Mg~+~NH collisions.
    The cross sections computed using the adiabatic method 
    are in excellent agreement with reference results obtained from the diabatic CC treatment \cite{Vieira:17,Morita:24,Wallis:09,GonzalezMartinez:11}, even near the scattering resonance shown in Fig.~\ref{fig:autotrunc_accuracy}(a).
This shows that  the fully converged adiabatic CC formulation  is as accurate as the rigorous diabatic CC approach \cite{Krems:04,Volpi:02}  when applied to  cold and ultracold atom-molecule collisions in a magnetic field. 

    We now turn to comparing the computational performance of the adiabatic basis to that of the standard diabatic basis. 
Because the computational time scales as the size of the basis cubed, $O(M^3)$, we compare the two approaches by reducing the number of channels used to propagate the log-derivative matrix and quantifying the error in the cross sections.
To do this, we solve the adiabatic eigenvalue problem, Eq.~\eqref{eq:adiabaticEigenvalueProblem}, at the midpoint of every sector ($R=c_n$) using the uncoupled-SF representation (Eq.~\eqref{eq:uncoupledSF}) with $\mathcal{N} = 954$ basis functions including all primitive states with $N \leq 6$, $L \leq 8$, and $M_{\text{tot}} = 1$. 
    We then reduce the number of adiabatic channels used to propagate the log-derivative matrix to a constant, $R$-independent value.
    For the diabatic basis, we simply reduce the number of diabatic channels, keeping it the same at all $R$, starting from the highest-energy, largest-$N$  basis states.

\begin{figure}[t!]
    \centering
    \includegraphics[width=0.45\textwidth]{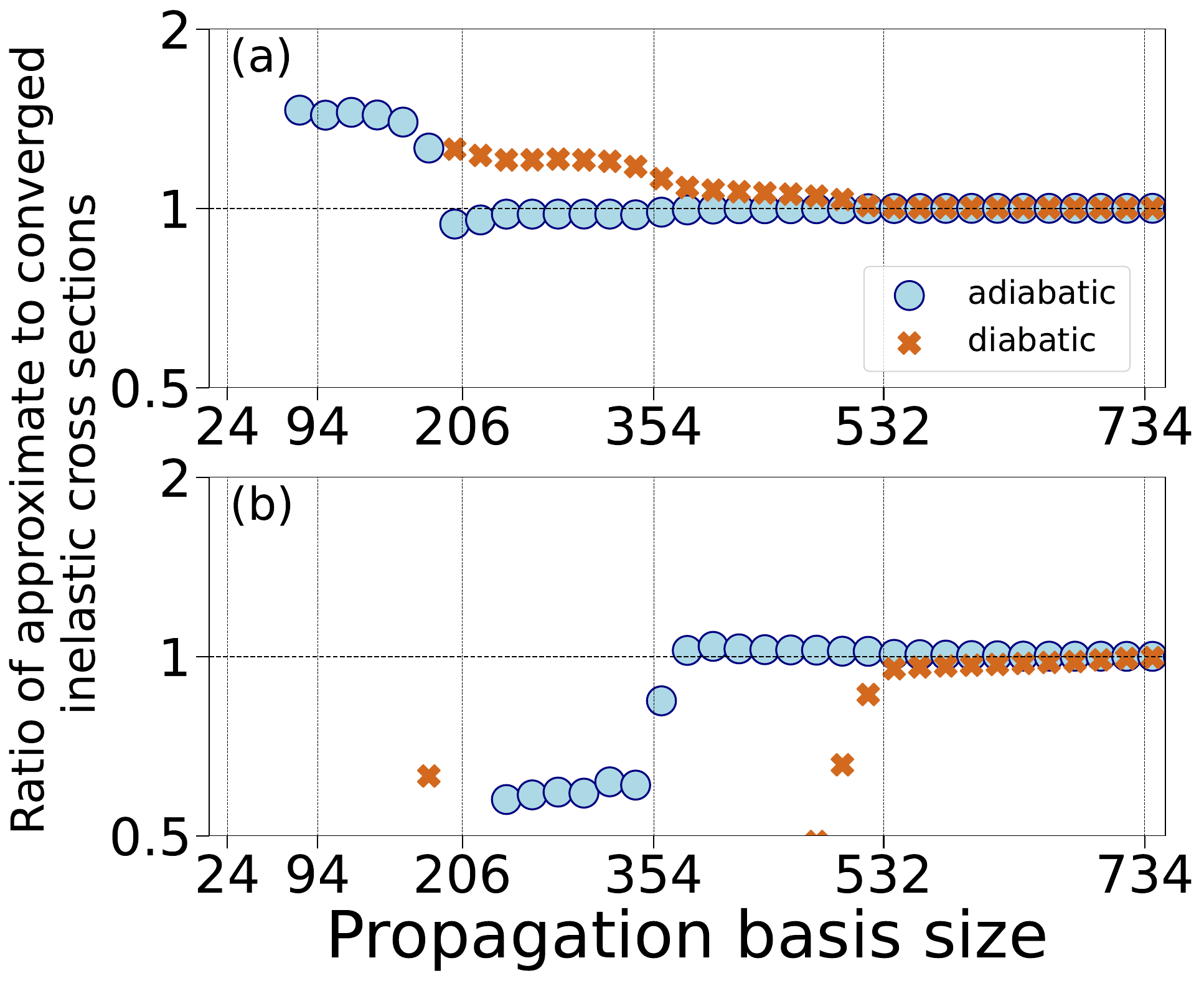}
    \caption{Ratios of the total inelastic cross sections calculated via a constant propagation basis size for the full propagation to the benchmark values computed with 954 basis states ($N_{\text{max}} = 6, L_{\text{max}} = 8$). Upper panel: Off-resonant collision energy $10^{-3} \text{cm}^{-1}$. Lower panel: Resonant collision energy $0.033 \text{cm}^{-1}$. Blue circles---adiabatic treatment, brown crosses---diabatic treatment. The vertical dashed lines represent cumulatively the number of states in each rotational manifold.}
    \label{fig:propagation_basis_convergence_mg-nh}
\end{figure}

Figure~\ref{fig:propagation_basis_convergence_mg-nh} shows the detailed effect of reducing the adiabatic basis size in an $R$-independent way. 
    To this end, we plot the ratios of the  calculated inelastic cross sections  to the benchmark values as a function of the basis size.
The top panel of Fig.~\ref{fig:propagation_basis_convergence_mg-nh} shows results for a non-resonant collision energy $E_\text{col} = 10^{-3} \text{ cm}^{-1}$, i.e., the energy away from the scattering resonances shown in Fig.~\ref{fig:autotrunc_accuracy}(a). The bottom panel of  Fig.~\ref{fig:propagation_basis_convergence_mg-nh} shows the results for the collision energy at the peak of the leftmost resonance feature shown in Fig.~\ref{fig:autotrunc_accuracy}(a), which occurs at $E_{\text{col}} = 0.033 \text{ cm}^{-1}$.
    We observe that at every basis size, the adiabatic basis produces cross sections with less error and provides high-quality approximate results with much fewer basis functions.
    Table~\ref{tab:coldtable} lists the number of basis functions required to compute the inelastic cross sections below a given percent error. 
The computational advantage of propagation in the adiabatic basis is approximated by cubing the ratio of the size of the diabatic basis ($M^{\text{di}}$) to the size of the adiabatic basis ($M^{\text{ad}}$).
    Table~\ref{tab:coldtable} shows that using the adiabatic basis results in the most significant advantage if larger percent errors can be tolerated.

\begin{center}
\begin{table}
\begin{tabular}{ | c | c c c | }
 \hline
  &  & $E_{col} = 10^{-3} \text{ cm}^{-1}$ & \\
 \hline
 \% error & $M^{\text{ad}}$ & $M^{\text{di}}$ & $\left( \frac{M^{\text{di}}}{M^{\text{ad}}} \right)^3$  \\
 \hline
 \hline
 0.1 & 540 & 720 & 2.4 \\ 
 1 & 380 & 540 & 2.9 \\ 
 5 & 220 & 480 & 10.4 \\ 
 10 & 200 & 380 & 6.9 \\ 
 50 & 80 & 200 & 15.6 \\ 
 \hline
\end{tabular}
    \caption{The basis size used for propagation required to compute inelastic cross sections at different levels of accuracy in the adiabatic ($M^{\text{ad}}$) and diabatic ($M^{\text{di}}$) bases. The basis size was kept constant for the propagation.}
    \label{tab:coldtable}
\end{table}
\end{center}

As shown in Fig.~\ref{fig:propagation_basis_convergence_mg-nh}(a), the diabatic basis provides results within $\simeq$20\% of the fully converged benchmark values with only three rotational manifolds in the basis ($N_{\text{max}}=2$), and that the adiabatic basis can really only push about one rotational manifold lower to get similar accuracy. 

Because the Mg-NH interaction is moderately anisotropic, 
    one may wonder whether the superior performance of the adiabatic basis will hold for deeper, more anisotropic potentials that require far more rotational states for convergence \cite{Tscherbul:11,Morita:18,moritaTscherbul2024}.
To test this hypothesis, we performed the same convergence test as that shown in Fig.~\ref{fig:propagation_basis_convergence_mg-nh} replacing the original Mg-NH PES with a much deeper and more anisotropic PES generated by  multiplying all Legendre expansion coefficients in Eq.~\eqref{eq:legendre_expansion} by a factor to 10. To scale up the relative anisotropy, we scaled the {\it anisotropic} Legendre expansion coefficients ($\lambda > 0$ in Eq.~\eqref{eq:legendre_expansion}) by a additional factor of two. The scaled potential $V'(R)$ can then be written as
            \begin{equation}
                V'(R,\theta) = 10 V_0(R) + 20\sum_{\lambda = 1}^{\lambda_{\text{max}}} V_{\lambda}(R) P_{\lambda}(\cos{\theta})
                \label{eq:scaled_PES}
            \end{equation}

\begin{figure}[h!]
    \centering
    \includegraphics[width=0.45\textwidth]{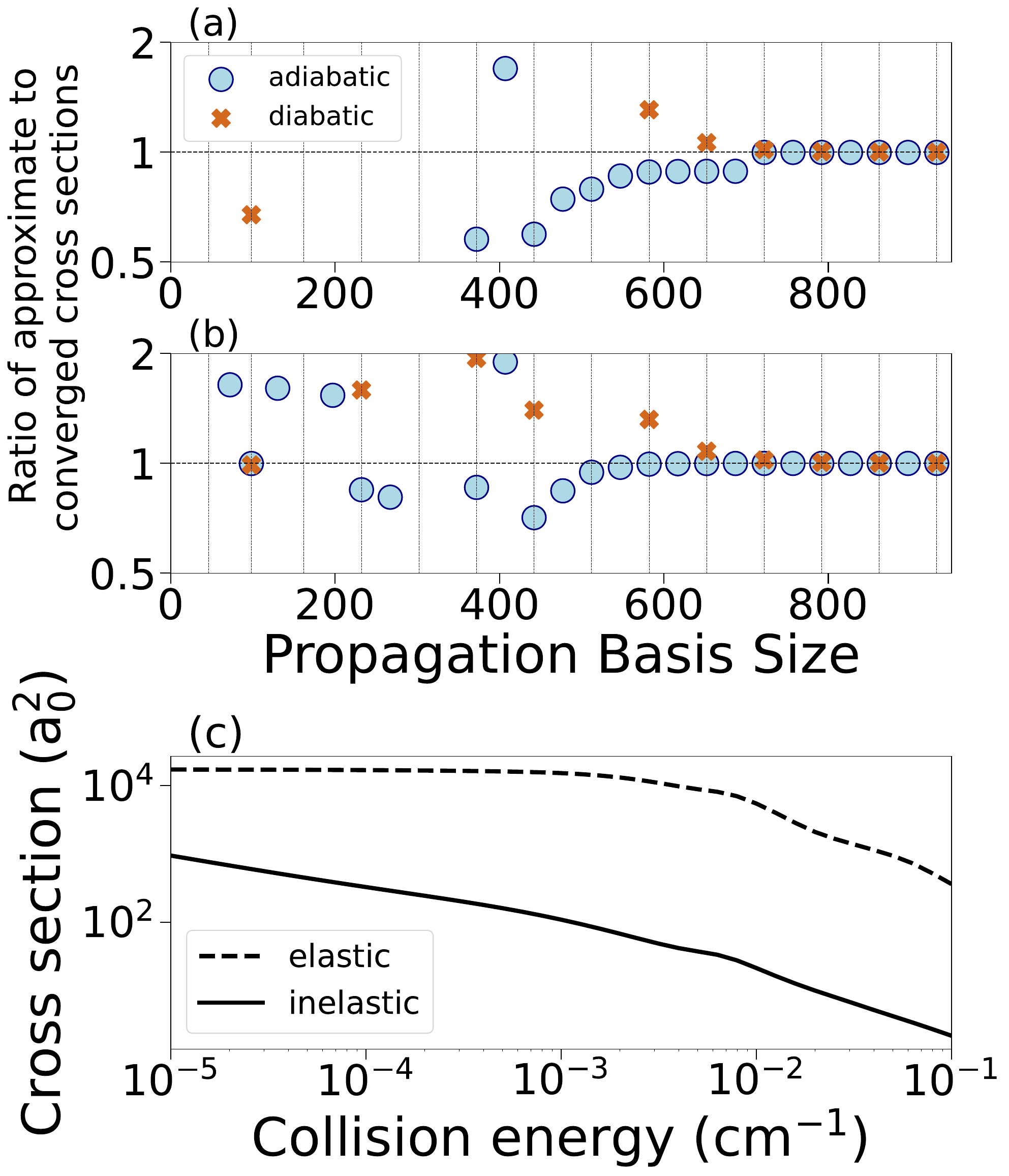}
    \caption{Same as Fig.~\ref{fig:propagation_basis_convergence_mg-nh} but for a PES that is scaled up as described by Eq.~\eqref{eq:scaled_PES}. The calculation was done in the TRAM basis. The benchmark values were obtained with 1002 basis states ($N_{\text{max}} = 15, J_{r,\text{max}} = 4$). 
    }
    \label{fig:propagation_basis_convergence_scaled}
\end{figure}

To accommodate such a deep and anisotropic PES, we solved the adiabatic eigenvalue problem  in the total {\it rotational} angular momentum (TRAM) representation as our primitive basis \cite{Simoni:06,tscherbulDincao2023,moritaTscherbul2024,Liu:25} (see Appendix~\ref{app:tram} and Ref.~\cite{tscherbulDincao2023} for computational details and matrix elements in the TRAM basis).
 Note that the uncoupled basis  cannot provide converged results for such highly anisotropic PESs \cite{Tscherbul:10,Tscherbul:11}.
  Approximately twice as many rotational manifolds are required to get converged cross sections using the scaled up PES ($N_{\text{max}} = 12$) compared to using the original Mg~+~NH PES ($N_{\text{max}} = 6$).
Figure~\ref{fig:propagation_basis_convergence_scaled} shows that the adiabatic basis requires 2-3 fewer rotational manifolds compared to the diabatic basis to compute cross sections that are within 20\% of the benchmark cross sections, resulting in a substantial reduction in the number of adiabatic channels even for the much deeper and more anisotropic PES. 
    This is in contrast to the 1-2 rotational manifolds that can be removed from the adiabatic basis when using the unscaled potential.

\begin{figure}[h!]
    \centering
    \includegraphics[width=0.45\textwidth]{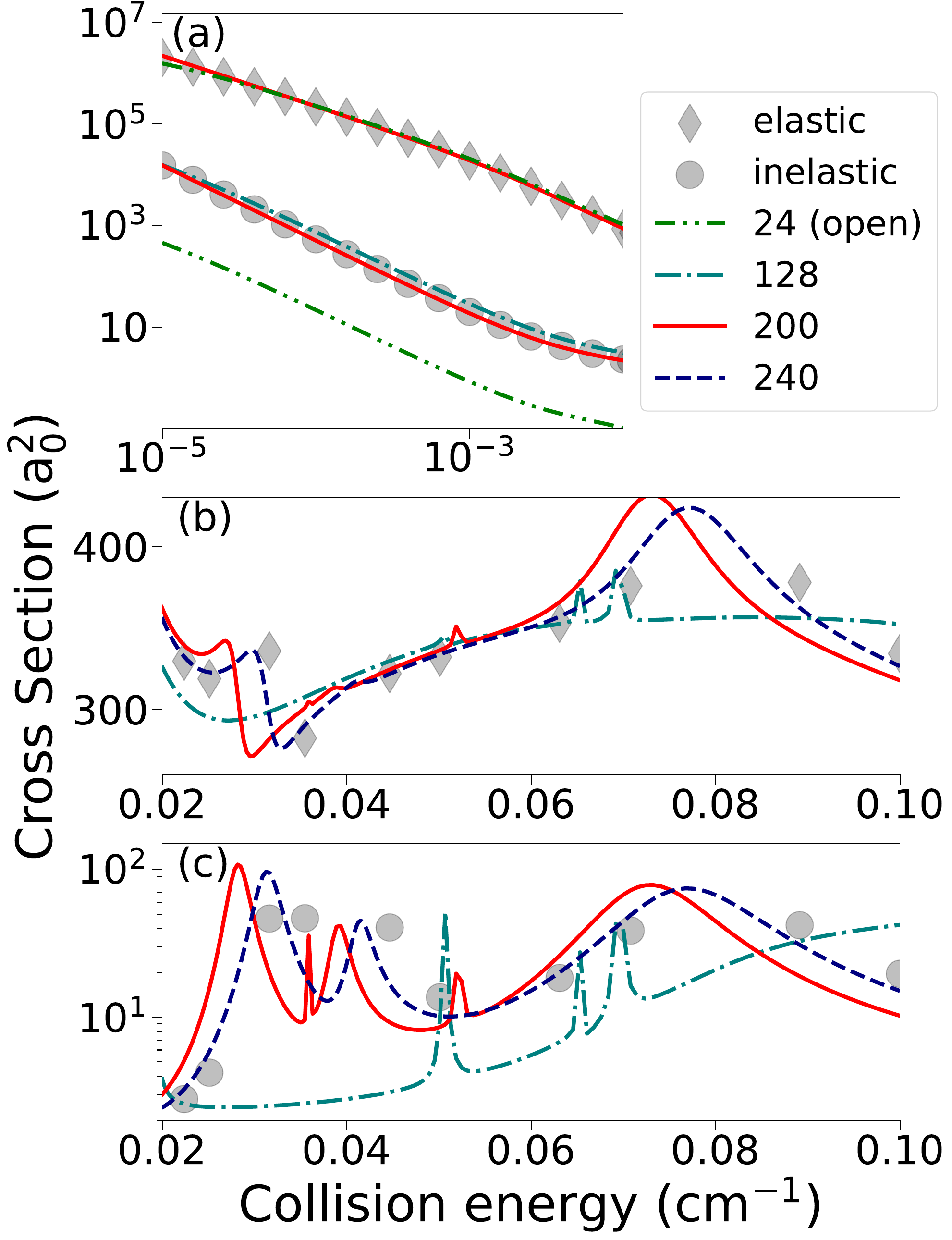}
    \caption{Cross sections of the Mg-NH collision vs. collision energy for different $R$-independent propagation basis sizes. All calculations were done without RBT. (a) Elastic and total inelastic cross sections at ultracold temperatures. (b) Elastic cross sections near resonances. (b) Total inelastic cross sections near resonances. All calculations were done at a magnetic field strength of $B = 100$~G. } 
    \label{fig:contrunc_accuracy_mg}
\end{figure}

Figure~\ref{fig:contrunc_accuracy_mg} shows the Mg~+~NH cross sections computed using the adiabatic basis of a fixed size $M$. 
    As for the data shown in Fig.~\ref{fig:propagation_basis_convergence_mg-nh}, the fully converged primitive basis was used to solve the adiabatic eigenvalue problem, and the $M$ channels with the lowest adiabatic eigenvalues were propagated.
At collision energies away from the resonance [see Fig.~\ref{fig:contrunc_accuracy_mg}(a)], the benchmark cross sections are well reproduced with just 200 adiabatic basis functions. 
    Reducing the number of adiabatic potentials to 128 channels (i.e., the 24 open channels and 104 weakly closed channels) does not strongly reduce the accuracy of the elastic cross sections, which are remarkably well reproduced with only the 24 open channels (see Fig.~\ref{fig:contrunc_accuracy_mg}) away from resonances.
However, the inelastic cross sections with just the open channels are underestimated by a factor of $\simeq$30. 

A total of 240 adiabatic basis functions are required to fully resolve the resonances with only a slight shift. 
    With 200 basis functions, there is a noticeable shift in the resonance positions and several spurious narrow resonances appear. 
With 128 adiabatic basis functions, the main resonance features are replaced by narrow resonances.
    This shows the importance of the strongly closed adiabatic channels in the vicinity of scattering resonances. 
Still, the Wigner s-wave limiting values of elastic (inelastic) cross sections can be reproduced with only 24 (128) adiabatic channels, a factor of 39.7 (7.5) smaller than the full number of adiabatic states ($\mathcal{N} = 954$). 

\begin{figure}
    \centering
    \includegraphics[width=0.65\textwidth]{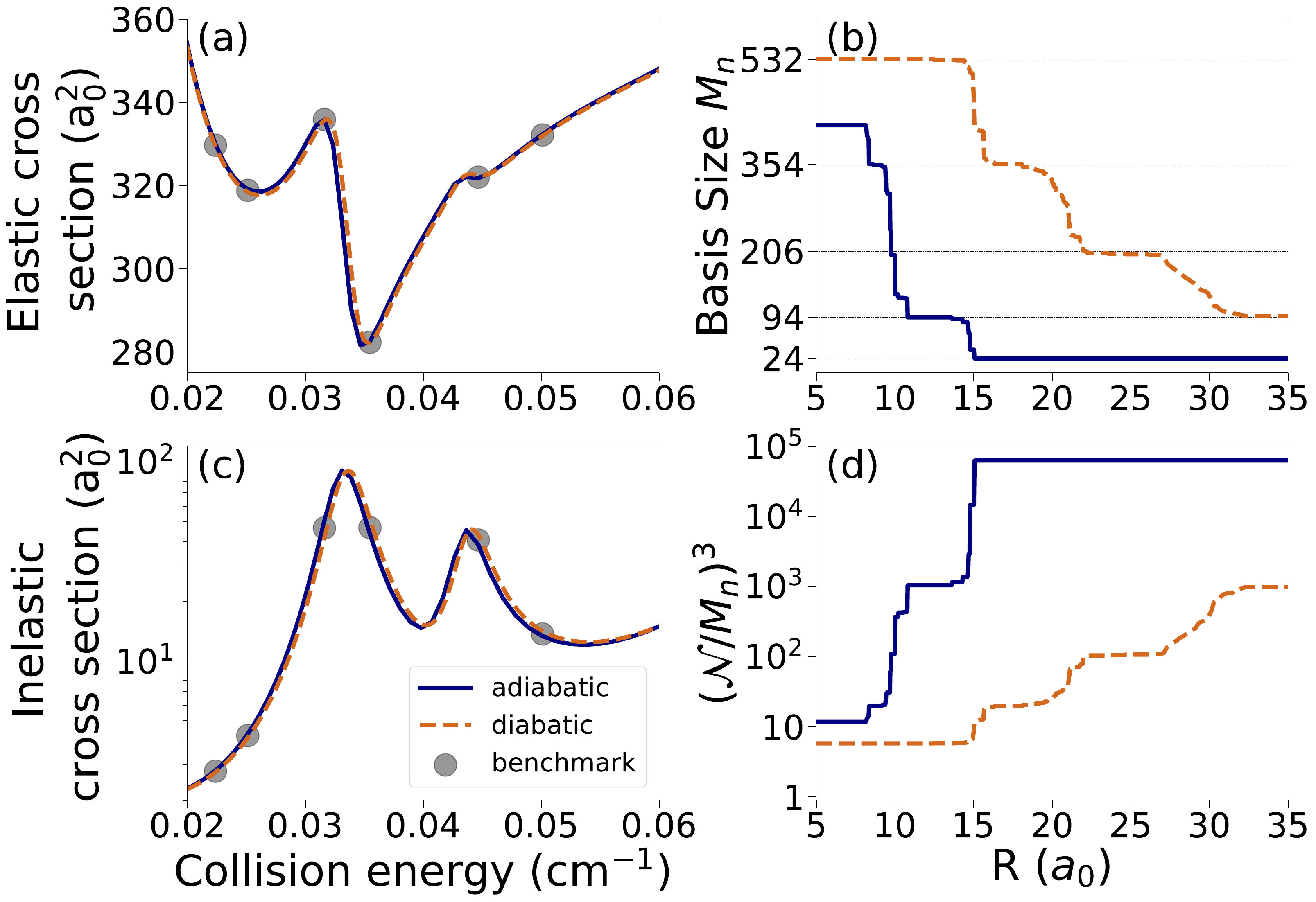}
    \caption{(a) Elastic and (c) total inelastic cross sections for cold Mg~+~NH collisions {in a magnetic field of strength $B = 100$~G} calculated using RBT with the $R$-dependent propagation basis size $M_n$ shown in panel (b). The RBT thresholds used for the adiabatic and diabatic treatments are $10^{-2} a_0^{-1}$ and $10^{-4} a_0^{-1}$, respectively. The initial propagation basis size before RBT is turned is $M_0=420$ for the adiabatic treatment and $M_0=532$ for the diabatic treatment. (d) Computational gain $( \mathcal{N} / M_n )^3$  of the adiabatic and diabatic RBT over the standard $R$-independent propagation  as a function of $R$, where $\mathcal{N}=954$ is the number of basis functions in the standard (fully uncoupled) basis.  The horizontal dashed lines in (b) represent cumulatively the number of states in each rotational manifold. The gray circles are benchmark results obtained via the full uncoupled space-fixed basis in the standard diabatic treatment.}
    \label{fig:comparison_resonance_exact_mg}
\end{figure}

\begin{figure}[t]
    \centering
    \includegraphics[width=0.65\textwidth]{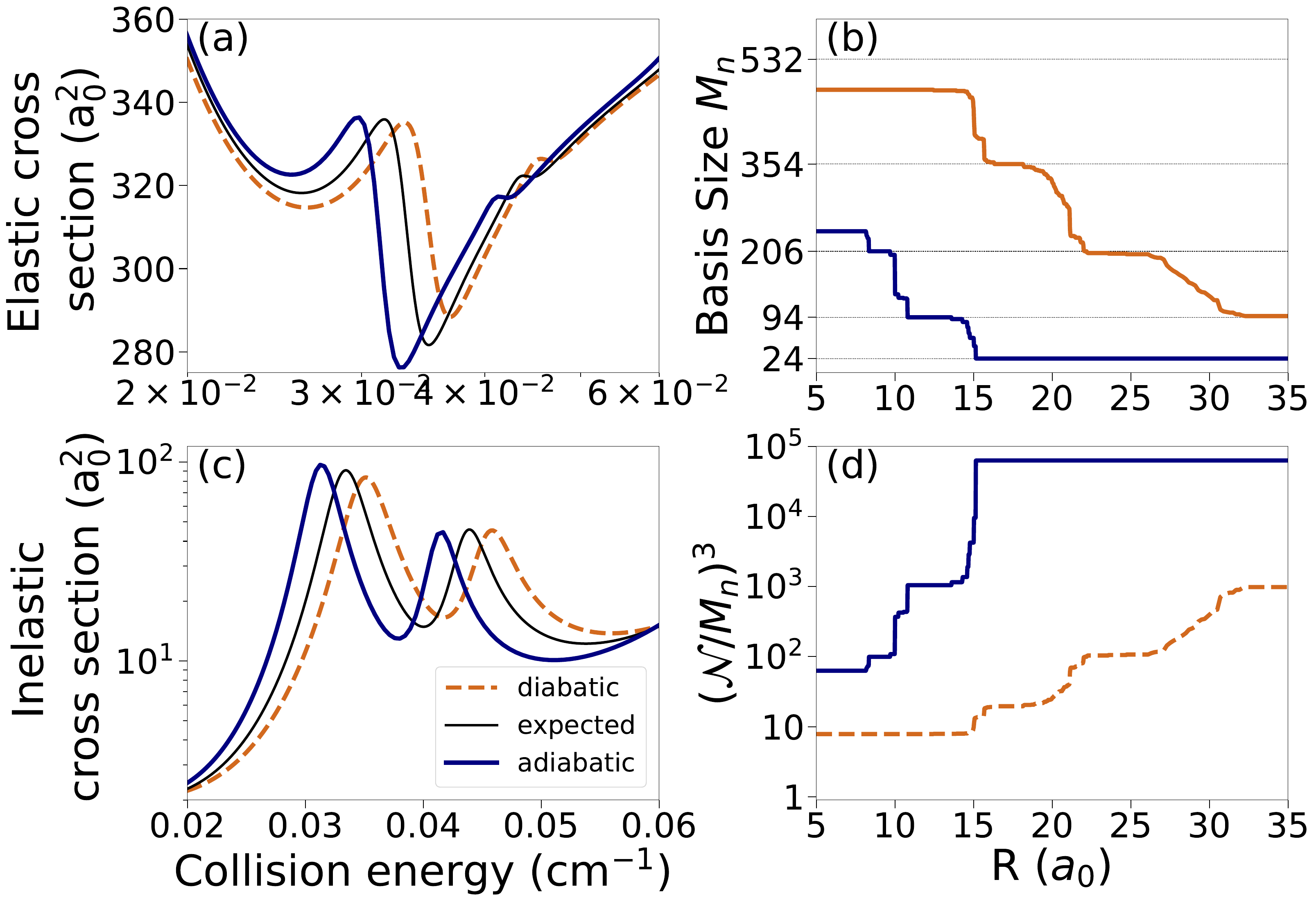}
    \caption{Same as Fig.~\ref{fig:comparison_resonance_exact_mg} with the same RBT thresholds but with initial basis sizes of $M_0=240$ and $M_0=480$ for the adiabatic and diabatic treatments, respectively.}
    \label{fig:comparison_resonance_approx_mg}
\end{figure}

\subsection{$R$-dependent basis truncation (RBT)}

\begin{figure}[t]
    \centering
    \includegraphics[width=0.65\textwidth]{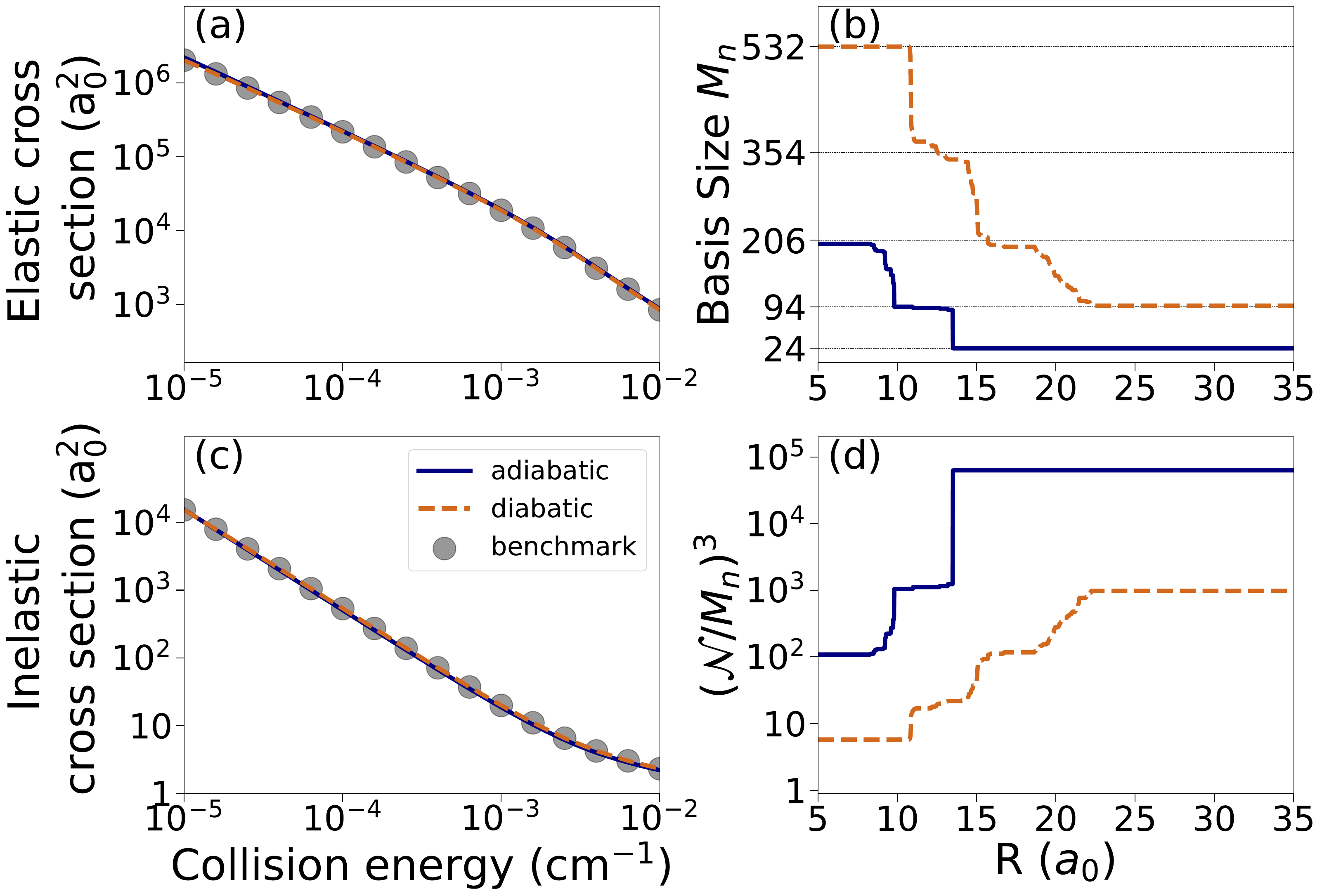}
    \caption{Same as Fig.~\ref{fig:comparison_resonance_exact_mg} but for ultracold collision energies and with RBT thresholds of $10^{-1} a_0^{-1}$ and $10^{-3} a_0^{-1}$ and initial basis sizes of $M_0=200$ and $M_0=532$ for the adiabatic and diabatic treatments, respectively.}
    \label{fig:comparison_cold_mg}
\end{figure}

Thus far we have used a fixed-size adiabatic basis at all $R$. We now invoke the RBT procedure described in Sec.~IIC and analyze its performance.
We set an RBT threshold that results in a small ($<5\%$) error in the cross sections. Another key RBT parameter is the initial propagation basis size used before RBT is turned on. 
We choose the initial basis size that produces nearly-exact results near the resonance (Fig.~\ref{fig:comparison_resonance_exact_mg}), approximate results near the resonance (Fig.~\ref{fig:comparison_resonance_approx_mg}), and nearly-exact results at ultracold collision energies (Fig.~\ref{fig:comparison_cold_mg}).

    The lower right panel of each figure shows the estimated computational gain offered by RBT over the standard CC calculation with $\mathcal{N}=954$ basis functions, given by $[ \mathcal{N}/M_n ]^3$. The computational gan increases gradually with $R$, as the number of propagated channels drops. The most significant  gains of 4-5 orders of magnitude  occur at $R\geq 15\,a_0$  in the abaibatic basis.

    These results show that the adiabatic RBT procedure is more efficient than the diabatic one at all $R$.
This is especially true at long range, where truncating to the open channels in adiabatic RBT is nearly 5 orders of magnitude  more computationally efficient than standard CC, and 2-3 orders of magnitude more efficient than diabatic RBT. 
    Because basis set truncation mostly occurs in the long-range region, it is the short-range region that dominates the computational time in RBT.
    
To  estimate the overall computational time for the propagation of the log-derivative matrix, we calculated the quantity
\begin{equation}
    \gamma = \sum_{n=1}^{N_S} M_n^3,
    \label{eq:computational estimate}
\end{equation}
where $N_S$ is the total number of sectors and $M_n$ is the number of basis functions used for the $n^{\text{th}}$ sector.
    We then take ratios of these sums to estimate the computational gains of adiabatic RBT ($\gamma_{a\text{RBT}}$) and diabatic RBT ($\gamma_{d\text{RBT}}$) over the benchmark fully converged CC calculation with 954 basis states ($\gamma_\text{954} = 954 N_S$).
The ratio $\gamma_\text{954}/\gamma_{a\text{RBT}}$ is 63.2, 306, and 539 for the basis size profiles from Figs.~\ref{fig:comparison_resonance_exact_mg}(b),\ref{fig:comparison_resonance_approx_mg}(b), and \ref{fig:comparison_cold_mg}(b), respectively. 
    Comparing adiabatic RBT to diabatic RBT, the ratios $\gamma_{d\text{RBT}}/\gamma_{a\text{RBT}}$ from the same three plots are 5.11, 18.3, and 27.4, respectively. 
    
    This suggests that adiabatic RBT offers the largest computational advantage when approximate cross sections (deviating slightly from the accurate resonance positions) can be tolerated, 
in keeping with the results from Table~\ref{tab:coldtable}. 
We note that these estimates  are only for the propagation part of the adiabatic CC calculation, and they do not account for the computational overhead 
 associated with solving the adiabatic eigenvalue problem.
    This is discussed further in Sec.~\ref{sec:conclusions}.
    
Diabatic RBT, however, {\it can} be directly compared to the benchmark CC calculation with 954 diabatic basis functions.
    The ratio $\gamma_\text{954}/\gamma_{d\text{RBT}}$ ranges from 12-20 for all three figures mentioned above, showing that diabatic RBT offers up to a 20-fold improvement in computational time when approximate cross sections can be tolerated.

\section{Summary and Conclusions} \label{sec:conclusions}

We have developed  a rigorous adiabatic approach to ultracold atom-molecule collisions in an external magnetic field.
    The approach is based on propagating the log-derivative matrix across a grid of sectors. 
     At each sector, one (1) solves the adiabatic eigenvalue problem at the midpoint of the sector, (2) propagates the log-derivative matrix across the sector, and (3) transforms the log-derivative matrix to the adiabatic basis of the next sector. 
Finally, at the end of the last sector, scattering boundary conditions are applied to the log-derivative matrix in the asymptotic basis. We applied the approach to ultracold Mg~+~NH collisions in a magnetic field, and found excellent agreement with standard CC calculations with a factor of $\simeq$2 reduction in the number of  scattering channels.
    We also developed an efficient log-derivative-based RBT method to reduce the size of the basis during propagation of the log-derivative matrix. 
    
As shown in Sec.~\ref{sec:results} for cold Mg~+~NH collisions, the adiabatic basis can be truncated significantly more aggressively than the diabatic basis, regardless of whether RBT is used, and  especially when an appreciable error ($<50$\%) in the calculated integral  cross sections can be tolerated. 
    That is, quality scattering observables can generally be obtained with far fewer basis functions in the adiabatic basis than the diabatic basis. 
In particular, by reducing the size of the basis without RBT, we found that the adiabatic approach can resolve the resonances near a collision energy of $0.04 \text{ cm}^{-1}$ with only 200 adiabatic basis functions, half of the basis functions required in the diabatic formulation. 
    Away from resonances, the adiabatic approach needs only the 128 open and weakly-closed channels to get accurate inelastic cross sections and only the 24 open channels for the elastic cross sections (Fig.~\ref{fig:contrunc_accuracy_mg}). 
By contrast, the diabatic approach needs at least 400 basis functions to obtain accurate cross sections either close to or away from resonances. 
    This gain is  expected because, unlike their diabatic counterparts, the adiabatic basis functions contain essential information about the scattering dynamics at fixed values of $R$.
Although we showed that this result holds for a deeper and more anisotropic version of the Mg~+~NH PES, it would be interesting to explore this for a realistic collision, such as Rb~+~SrF \cite{moritaTscherbul2024}.
    
Comparing the computational performance of the adiabatic vs. diabatic basis sets is challenging because the two approaches differ in the number of floating-point operations required per sector at the same basis size. 
    We estimate that the adiabatic approach as presented here requires approximately 15 times more floating-point operations for the propagation part of the calculation  than the standard diabatic method at the same basis size (see Appendix~\ref{app:compdetails} for details).
     Considering this extra computational cost and using Eq.~\eqref{eq:computational estimate} and the results in Figs.~\ref{fig:comparison_resonance_approx_mg} and \ref{fig:comparison_cold_mg}, we estimate the adiabatic basis is 15-30 times more computationally efficient than the standard diabatic basis with 954 basis functions, provided small shifts in resonance positions can be tolerated.
Using the same metrics, the computational performance of the adiabatic RBT is on par with that of the diabatic RBT.   

  Our results show that the adiabatic approach provides the largest computational advantage at long range because the adiabatic basis can be truncated down to just the open channels already at $R\ge 15\,a_0$.
Recent quantum reactive scattering calculations in the absence of external fields also reported  a notable reduction in the number of adiabatic states $(860\to 500)$ when  projecting the wavefunction of the LiNa$_2$ reaction complex from adiabatically adjusting principal axis hyperspherical (APH) to Delves coordinates  \cite{Kendrick:21}. The reduction observed in Ref.~\cite{Kendrick:21} is, however,  much less dramatic than that reported here ($420 \to 24$), likely due to a large number of open channels in the ultracold Li~+~NaLi $\to$ Na + Li$_2$ reaction.

In notable contrast to the adiabatic basis, the diabatic basis does not produce accurate cross sections if truncated below about 90 basis functions due to the coupling between rotational states with different $N$ in the asymptotic Hamiltonian (see Appendix~\ref{app:diabatic_RBT}).
The ability to use only the open channels to propagate the log-derivative matrix at long range is an attractive feature of the adiabatic approach
because the number of open channels is generally much smaller than the total number of channels, and does not depend on the atom-molecule interaction anisotropy (or  $N_{\text{max}}$).
    This also suggests that  the adiabatic approach could be most advantageous for ultracold molecular collisions mediated by long-range interactions, such as NH~+~NH \cite{Suleimanov:12,Suleimanov:16}, where the log-derivative matrix must be propagated out to very large atom-molecule separations ($R_{\text{max}} = 500 \text{ a}_0$ \cite{Suleimanov:12}).

The concept of universality in ultracold collision physics states that all of (or most of) the physics can be encapsulated in a small number of short-range parameters \cite{Burke_98,Gao_05,Croft:11,Morita:24} that are independent of  either the collision energy or magnetic field. These universal parameters   are typically described in the framework of multichannel quantum defect theory, which leverages the separation of  energy scales in ultracold collisions to arrive at a simplified description of two-body collision physics \cite{Morita:24}  in terms of a few short-range parameters. However,  this requires one to assume that some  degrees of freedom (such as nuclear spins) play a spectator role.
The recent experimental observation of hyperfine-to-rotational energy transfer  in ultracold Rb~+~KRb collisions suggests that  short-range couplings between the spin and rotational degrees of freedom can be non-negligible, posing a challenge for MQDT-FT \cite{Morita:24}. 

Our RBT approach does not rely on  approximations and systematically truncates the log-derivative matrix as it is propagated from small to large $R$. By stopping the propagation at an intermediate value of $R_\text{open}$, where the log-derivative matrix is maximally truncated to the open-channel basis, one obtains the log-derivative matrix $\mathbf{Y}(R_\text{open})$, which contains all information about scattering observables. This provides an alternative, numerically exact way of  condensing the complex physics of multichannel atom-molecule collisions into a small number of short-range  parameters, the matrix elements of $\mathbf{Y}(R_\text{open})$. For these parameters to be maximally useful, they should be independent of the collision energy and external fields. Whether or not this is the case remains to be explored.
It would also be interesting to extend the efficient basis sets for solving the adiabatic eigenvalue problem (such as the TRAM representation \cite{tscherbulDincao2023}) to include additional degrees of freedom, such as molecular vibrations and multiple potential energy surfaces. 

\begin{acknowledgments}
We are grateful to Chris Greene and Jose P. D'Incao for stimulating discussions.
This work was supported by the NSF CAREER award No. PHY-2045681.
\end{acknowledgments}

\appendix

\section{Computational details} \label{app:compdetails}

In all calculations, we propagate the log-derivative matrix from $R = 4 \text{ } a_0$ to $R = 25 \text{ } a_0$ with a step size of $\Delta R_1 = 0.01 \text{ } a_0$.
    We then propagate to $R = 100 \text{ } a_0$, where we apply $S$-matrix boundary conditions. 
The step size for propagating the log-derivative matrix for $R > 25 \text{ } a_0$ is $\Delta R_2 = 0.1 \text{ } a_0$.
The values of parameters used in our CC calculations are listed in Table~\ref{tab:interaction_parameters}.
We employ the same 954-channel uncoupled SF basis given by Eq.~\eqref{eq:uncoupledSF} in benchmark diabatic CC calculations, as well as to solve the adiabatic eigenvalue problem \eqref{eq:adiabaticEigenvalueProblem}.
This basis includes  all states with $N \leq 6$, $L \leq 8$, and $M_{\text{tot}} = 1$ and gives fully converged results.
We choose the upper Zeeman state of the $N=0$ rotational manifold as our initial state, and perform calculations for the $M_{tot} = M_N + M_L + M_S = 1$ block of states.


\begin{center}
\begin{table}[b]
   \caption{Values of parameters used in our Mg~+~NH scattering calculations.}
\begin{tabular}{  c  c  }
 \hline
  \hline
Parameter & Value \\
 \hline
Rotational contant of NH, $B_e$ & 16.32176 cm$^{-1}$ \\
Spin-rotation constant of NH, $\gamma_{\text{sr}}$ & -0.05467 cm$^{-1}$  \\
 Spin-spin constant of NH, $\lambda_{\text{SS}}$ & 0.9197 cm$^{-1}$ \\ 
 Reduced mass of Mg-NH, $\mu_{\text{Mg-NH}}$ & 9.23267993 amu \\
 \hline
\end{tabular}
    \label{tab:interaction_parameters}
\end{table}
\end{center}

\begin{figure}[t!]
    \centering
    \includegraphics[width=0.55\textwidth]{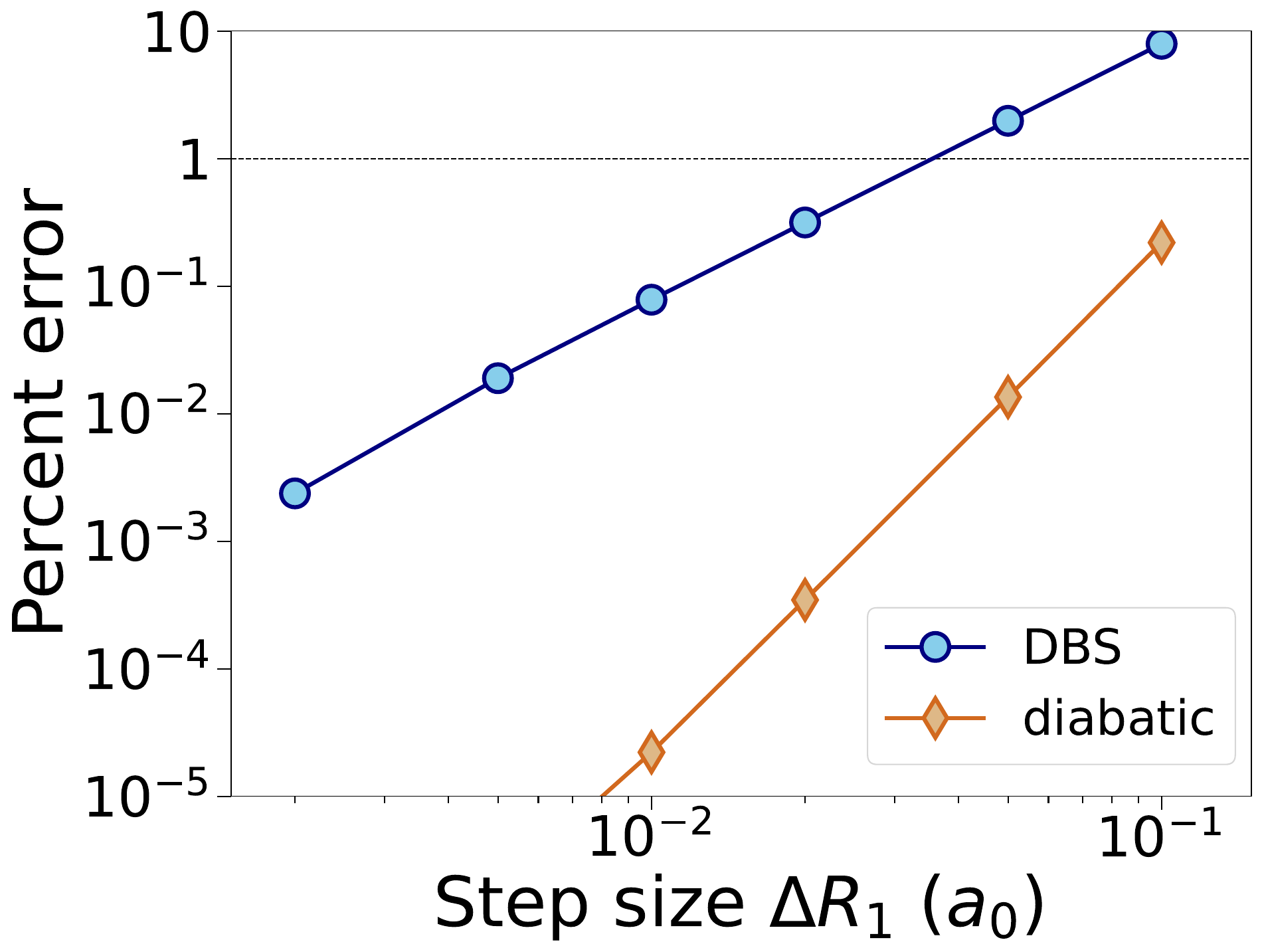}
    \caption{Percent error in the inelastic cross section plotted as a function of the step size $\Delta R_1$ with $\Delta R_2$ kept fixed (see text).
    The total inelastic cross section with a grid spacing of $\Delta R_1 = 0.001 \text{ } a_0$ is used as the benchmark value in computing the error values.}
    \label{fig:dR_convergence}
\end{figure}

The matrix operations involved in the propagation  part of the calculation scale as the size of the basis cubed.
There are  two matrix inversions  in the diabatic approach, whereas the adiabatic basis requires an orthogonal transformation in addition to the two inversions.
    Symmetric matrix inversion scales as $\simeq M^3$ and orthogonal transformations as  $\simeq 3M^3$ \cite{Trefethen:97}.
From these rough estimates and a simple speed test, we find that it takes roughly two times more floating-point operations to propagate the log-derivative across a sector in the adiabatic basis than in the diabatic basis. 
In addition, we found that the diabatic-by-sector approach requires about 5-10 times more sectors than the diabatic method (Fig.~\ref{fig:dR_convergence}). 
    This amounts to an overall factor of $\simeq 10{-}20$ increase in the computational time required to propagate the log-derivative matrix in the adiabatic basis compared to the diabatic basis. 
We take the midpoint of this range, and conclude that the adiabatic approach requires $\simeq  15$ times more floating-point operations to propagate the log-derivative through a range of $R$-values for the same basis size compared to the diabatic approach. Fortunately, this increase is more than offset by the more efficient performance of RBT in the adiabatic basis, as described below.

\section{Truncating the basis during propagation} \label{app:RBT}

\begin{figure}[h!]
    \centering
    \includegraphics[width=0.50\textwidth]{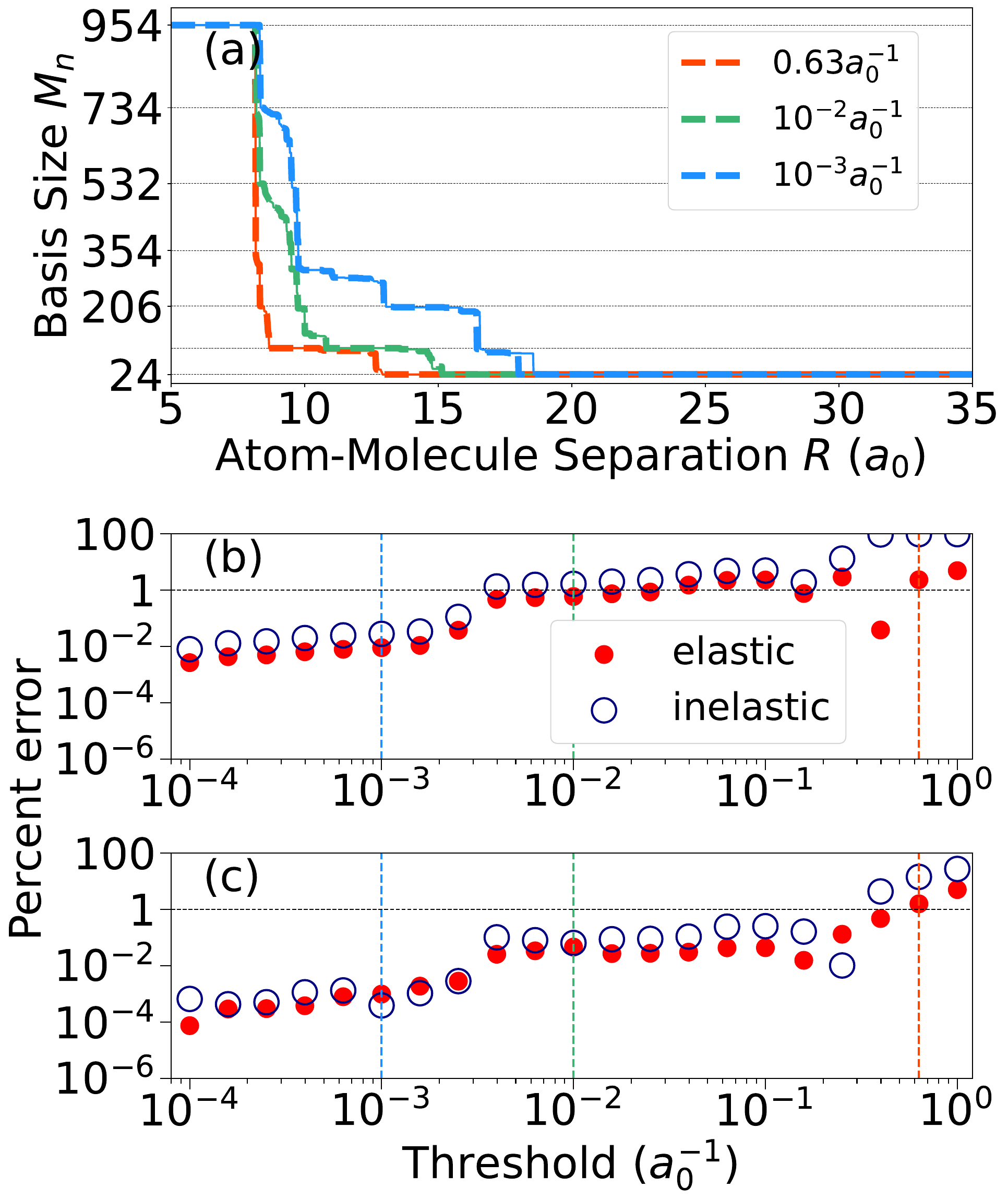}
    \caption{Log-derivative-based RBT in adiabatic basis. (a) Basis size $M_n$ against atom-molecule separation $R$ for three different RBT thresholds. The horizontal dashed lines indicate the cumulative number of channels in each rotational manifold up to $N = 6$. (b) Percent error in the elastic (filled red circles) and total inelastic (open blue circles) cross sections against RBT threshold at $E_{\text{col}} = 0.033 \text{ cm}^{-1}$ and $B = 100 \text{ G}$. The vertical dashed lines show the RBT thresholds for which we plot the basis size profiles in (a). The horizontal dashed line marks $1\%$ error as a guide to the eye. (c) Same as (b) but for $E_{\text{col}} = 10^{-3} \text{ cm}^-1$. }
    \label{fig:adiabatic_rbt}
\end{figure}

Figure~\ref{fig:adiabatic_rbt}(a) shows the number of basis states used for propagation as a function of $R$, for log-derivative-based RBT in the adiabatic basis. 
    These RBT ``trajectories,'' { or basis size profiles,} are plotted for different RBT thresholds that produce cross sections of varying accuracy. 
The corresponding errors introduced by RBT are shown in Figs.~\ref{fig:adiabatic_rbt}(b) and \ref{fig:adiabatic_rbt}(c) for resonant and non-resonant collision energies, respectively.
    Analyzing the RBT errors  gives insight into which adiabatic states are important in different regions of $R$. 
Compare, for example, the basis size profile for $\tau_{\text{RBT}} = 10^{-2} \text{ } a_0^{-1}$ (green curve), which introduces a very small ($< 0.1\%$) error with that  for $\tau_{\text{RBT}} = 0.63 \text{ } a_0^{-1}$ (orange curve), for which  the error is large ($\simeq 100\%$).
    These two basis size profiles mainly differ in that the green curve retains $\approx 400$ adiabatic channels until $R=10 \text{ } a_0$ and the adiabatic channels in the $N=1$ manifold until $R = 15 \text{ } a_0$.
The orange curve truncates the adiabatic basis to just the channels in the $N = 0$ and $N=1$ manifolds by $R \approx 8 \text{ } a_0$, and just the open channels by $R \approx 13 \text{ } a_0$. 
    {\it The significant difference in accuracy of cross sections resulting these two basis size profiles indicates the importance of the additional adiabatic channels retained in the basis for $\tau_{\text{RBT}} = 10^{-2} \text{ } a_0$ in the two regions of  $R$ discussed above.}

\begin{figure}[t!]
    \centering
    \includegraphics[width=0.55\textwidth]{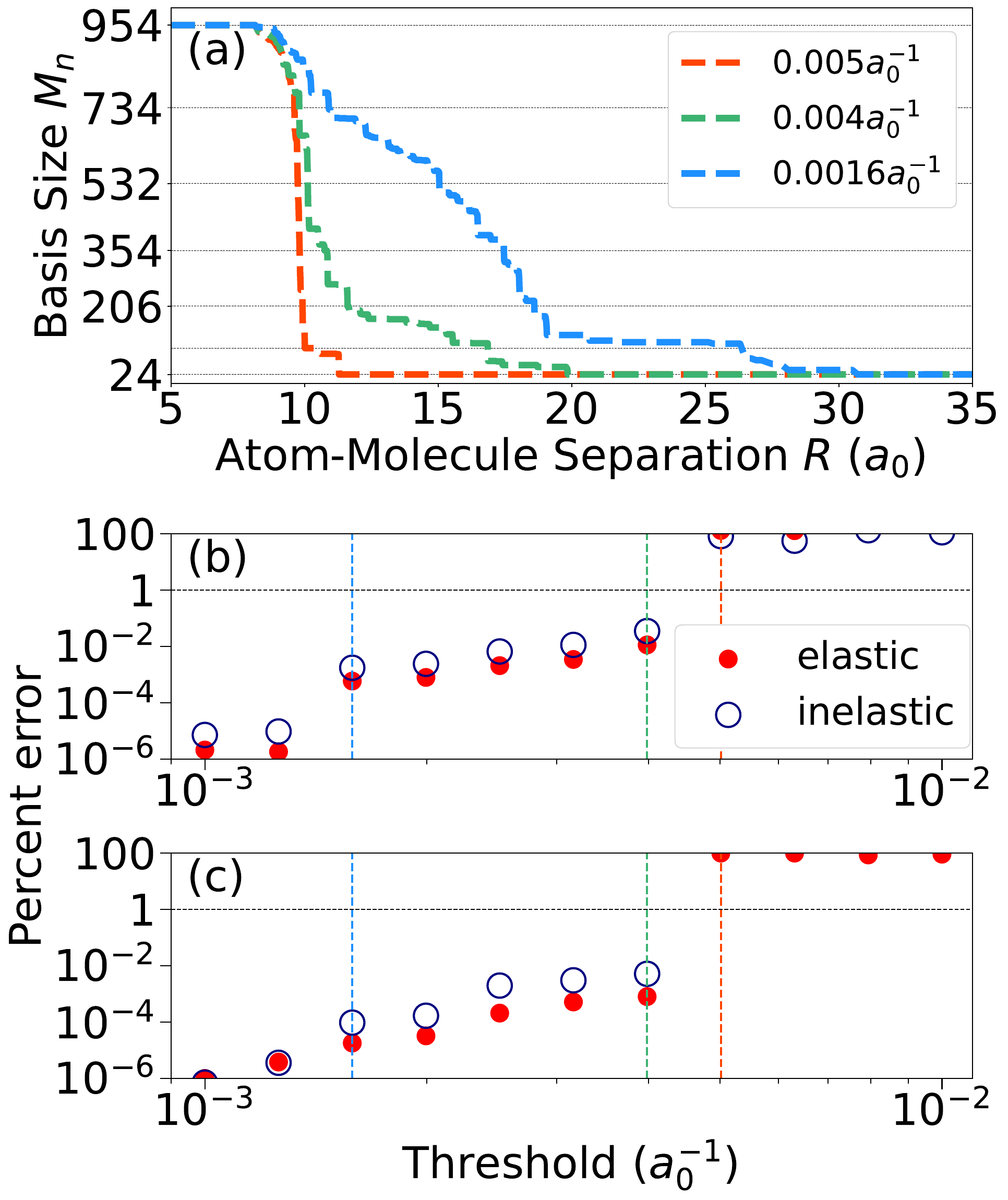}
    \caption{Same as Fig.~\ref{fig:adiabatic_rbt} but for overlap-based RBT, which is necessarily in the adiabatic basis.}
    \label{fig:overlap_rbt}
\end{figure}

We also tested the performance of overlap-based RBT, in which the matrix elements of the overlap matrix (Eq.~\eqref{eq:overlap}) are sampled to determine the channel to be removed. 
    We found the best implementation to be close to that proposed by Stetchel, Walker, and Light \cite{Stetchel:1978}.
They suggest removing the $\alpha^{\text{th}}$ channel if the quantity
        \begin{equation}
            \beta_{\alpha} = \frac{1}{2(\alpha-1)\Delta R} \left[ \sum_{j=1}^{\alpha - 1} \left( {\bf O}_{j \alpha}^2 + {\bf O}_{\alpha j}^2 \right) \right]^{\frac{1}{2}}
            \label{eq:orbt_condition}
        \end{equation}
is less than the user-defined RBT threshold $\tau_{\text{RBT}}$, where $\alpha$ is the index of the channel being tested and ${\bf O}_{j\alpha}$ are matrix elements of the overlap matrix (Eq.~\eqref{eq:overlap}).
    As in log-derivative-based RBT in the adiabatic basis, we start the truncation from the adiabatic channels with the highest energy, and continue in the order of decreasing energy.
We stop truncating once we find the first state that does not satisfy $\beta_{\alpha} < \tau_{\text{RBT}}$.
    Figure~\ref{fig:overlap_rbt} shows the {basis size profiles and errors in inelastic cross sections} with respect to $\tau_{\text{RBT}}$ for overlap-based RBT.
An interesting aspect of overlap-based RBT is that the error in the cross sections changes rapidly from $\approx 10^{-2} \%$ to $\approx 100 \%$  when $\tau_{\text{RBT}}$ is varied slightly from $\tau_{\text{RBT}} = 0.004 \text{ } a_0^{-1}$ to $0.005 \text{ } a_0^{-1}$.
    Comparing the basis size profiles in Fig.~\ref{fig:adiabatic_rbt}(a) and Fig.~\ref{fig:overlap_rbt}(a) for similar levels of accuracy, we find that log-derivative-based RBT produces basis size profiles that deviate only slightly from those of overlap-based RBT.
We note that basis set profiles in overlap-based RBT can be obtained without propagating the log-derivative matrix.

The solid lines in Fig.~\ref{fig:adiabatic_rbt}(a) differ from the dashed lines in that the basis size profiles were produced by sampling the log-derivative matrix at a different collision energy.
    Comparing these shows that the basis size profiles are largely independent of collision energy.
This makes implementation easier because the log-derivative matrix needs only to be sampled at a single collision energy, and the resulting basis size profiles can be used for other collision energies (this was only tested for collision energies below $0.1 \text{ cm}^{-1}$).
{The overlap matrix is independent of the collision energy, and so are overlap-based RBT basis size profiles.}

\subsection*{Diabatic RBT} \label{app:diabatic_RBT}

A major advantage of log-derivative-based RBT over overlap-based RBT is that the former can be applied in the diabatic basis.
    Below we describe the main features of RBT in the diabatic basis (diabatic RBT),  including the procedures for  (i) how to define the locally open and locally closed channels, (ii) how to search for the states to be removed, and (iii) when to stop the search.
    
  First, in diabatic RBT, we label a basis state as locally open or locally closed based on the diagonal matrix element of the reference potential, defined by Eq.~\eqref{eq:reference_potential} with the adiabatic eigenvector matrix ${\bf T}_n$ replaced by the identity matrix.
The $i^{\text{th}}$ primitive basis state is locally open if $[{\bf W}_{\text{ref}}]_{ii} < 0$ and locally closed if $[{\bf W}_{\text{ref}}]_{ii} > 0$.
    
\begin{figure}[t!]
    \centering
    \includegraphics[width=0.55\textwidth]{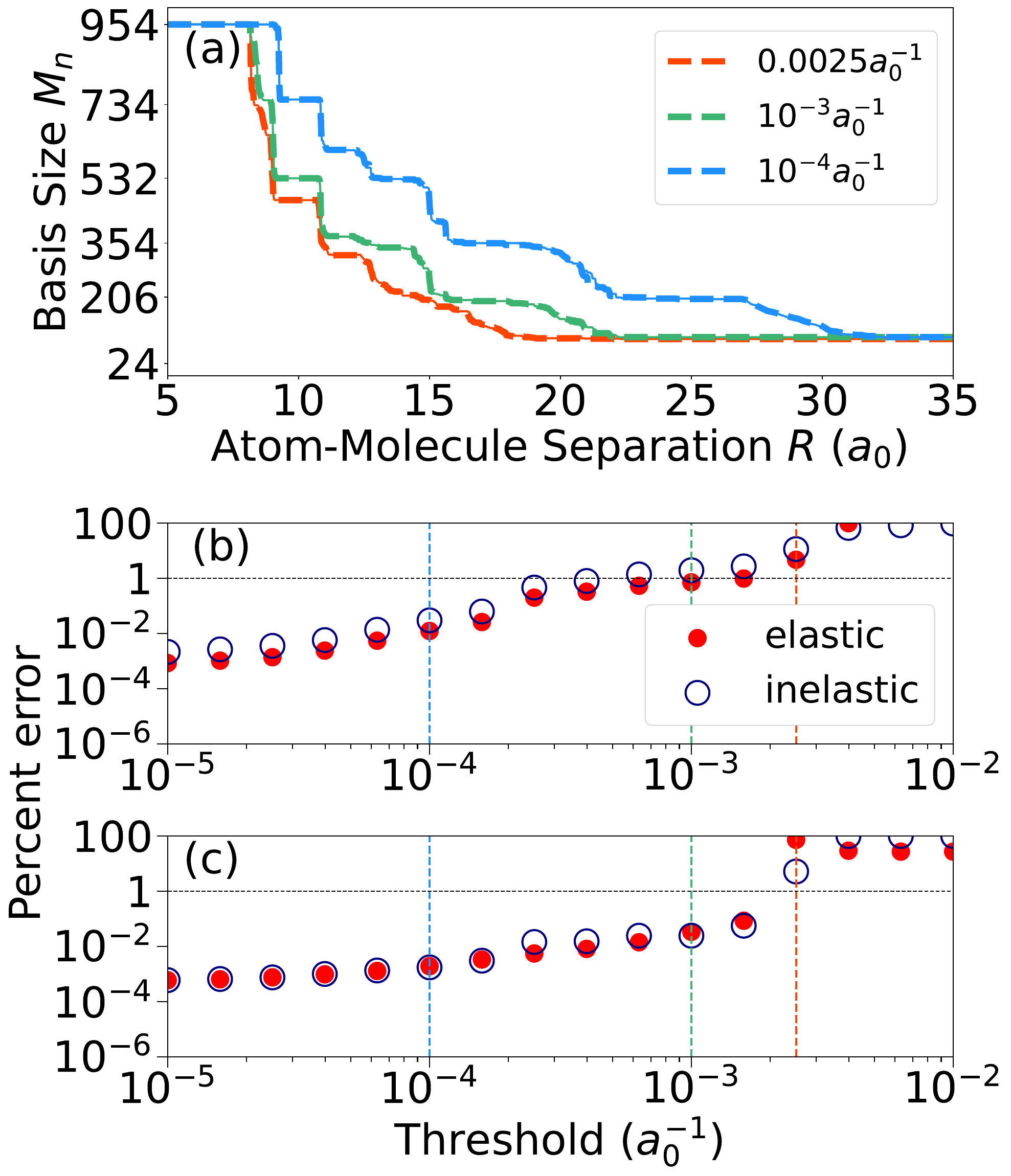}
    \caption{Same as Fig.~\ref{fig:adiabatic_rbt} but for  diabatic RBT.}
    \label{fig:diabatic_rbt}
\end{figure}

The performance of diabatic RBT is illustrated in Fig.~\ref{fig:diabatic_rbt}.
    As expected, the error in the cross sections steadily decreases as the RBT threshold is decreased. 
For a given accuracy, the basis size obtained with diabatic RBT is larger than that with adiabatic RBT at almost all $R$ values. 
    This is consistent with the results shown in Figs.~\ref{fig:comparison_resonance_exact_mg}-\ref{fig:comparison_cold_mg} in the main text. 
We can also quantify the computational gain provided by diabatic RBT by using Eq.~\eqref{eq:computational estimate} and calculating the ratio $\gamma_{954}/\gamma_{d\text{RBT}}$.
    From these ratios,  the  green and blue basis size profiles in Fig.~\ref{fig:diabatic_rbt} (both of which provide accurate cross sections) correspond to 4- and 5.5-fold computational gains over the standard diabatic CC basis consisting of 954 functions.

As noted above, in adiabatic RBT we start channel elimination from the highest-energy adiabatic state in each sector and continue until we encounter a channel that does not meet the cutoff condition. The entire procedure is then stopped, a procedure we refer to as the ``top-down approach''.
In diabatic RBT, an alternative ``any-state'' approach is possible, where {\it any} locally closed channel that satisfies the cutoff condition can be removed, regardless whether or not all the higher-energy channels have already been eliminated. 
The any-state approach is thus expected to trim the diabatic basis  more aggressively as a function of $R$.


\begin{figure}[t!]
    \centering
    \includegraphics[width=0.65\textwidth]{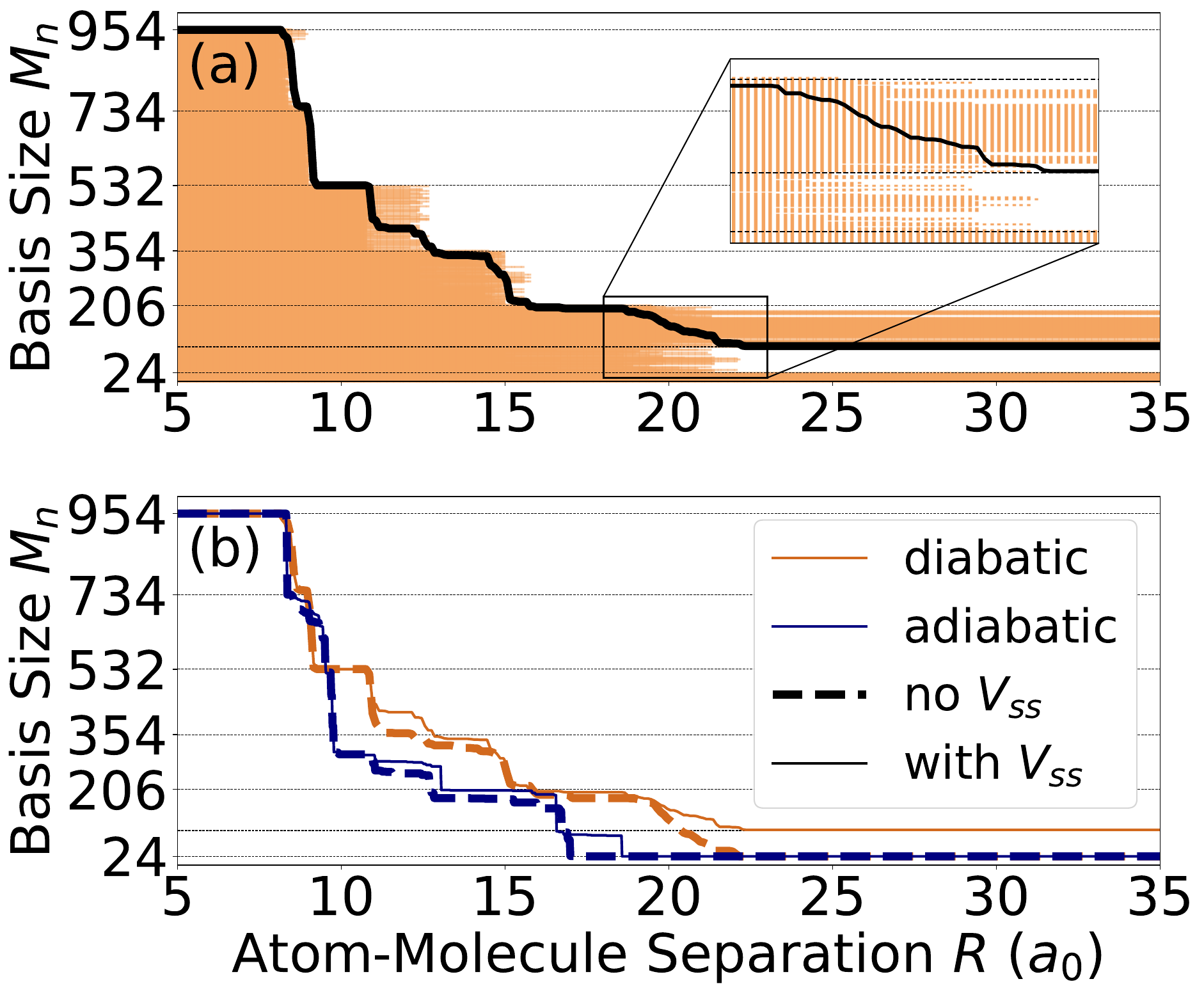}
    \caption{(a) Basis size profile (black line) for diabatic RBT at $E_{\text{col}} = 10^{-3} \text{ cm}^{-1}$, {$B = 100$~G}, and an RBT threshold of $10^{-3} \text{ } a_0^{-1}$. At each $R$-value, the brown points mark the indices of the diabatic basis functions retained in the basis. The inset shows the region $R \in [18, 23] \text{ } a_0$ to highlight the truncation of the diabatic basis functions with $N=1$ while retaining the functions with $N=2$. (b) The basis size profiles for adiabatic (blue) and diabatic (brown) RBT at $E_{\text{col}} = 10^{-3} \text{ cm}^{-1}$ and an RBT threshold of $10^{-3} \text{ } a_0^{-1}$. The dashed lines correspond to calculations with the spin-spin interaction in the asymptotic Hamiltonian turned off and the solid lines correspond to full CC results with the spin-spin interaction included. }
    \label{fig:noVss}
\end{figure}

    Figure~\ref{fig:noVss}(a) shows the basis size profile for any-state diabatic RBT.
Also plotted are markers (orange points) showing the indices of diabatic basis functions retained in the basis. 
    The inset highlights how any-state  diabatic RBT is able to remove all of the $N=1$ basis functions while keeping the $N=2$ functions, resulting in just 96 basis functions. This is a factor of 2 fewer channels than with the top-down approach, which truncates to $\simeq$190 basis functions, retaining almost all states up to and including $N=2$ functions.

The diabatic basis must retain the $N=2$ primitive basis functions at long range because of the spin-spin interaction in the asymptotic Hamiltonian \eqref{eq:asymptotic_hamiltonian}, 
    which couples  the $N=0$ states that dominate the open channels to the $N=2$ primitive states \cite{Krems:04}.
Figure~\ref{fig:noVss}(b) compares the basis size profiles with the spin-spin interaction included (thin solid curves) vs omitted  (thick dashed curves). 
    A significant difference occurs at $R = 21 \text{ } a_0$, where the basis size profile with the spin-spin interaction omitted truncates to just the 24 open channels, whereas the full calculation does not truncate below 96 diabatic channels. 
The basis size profile for adiabatic RBT is not significantly affected by removing  the spin-spin interaction because this coupling is already incorporated in the adiabatic basis functions.

\section{Matrix elements of the spin-spin interaction in the TRAM basis} \label{app:tram}

Here, we provide the expressions for the matrix elements of the adiabatic Hamiltonian in the TRAM basis.
    The TRAM is a vector sum of angular momenta for mechanical rotation, $\hat{{\bf J}}_r = \hat{{\bf N}} + \hat{{\bf L}}$.
The TRAM basis functions used as our primitive basis are eigenstates of $\hat{J}_r^2$, $\hat{J}_z$, $\hat{N}^2$, and $\hat{L}^2$, and can be written in terms on the uncoupled states $\ket{N M_N}\ket{L M_L}$ as
\begin{equation}
    \ket{(N L) J_r M_r} = \sum_{M_N,M_L} \langle N M_N L M_L | J_r M_r \rangle \ket{N M_N}\ket{L M_L}.
    \label{eq:tram_expansion}
\end{equation}
The matrix elements for all interactions in the adiabatic Hamiltonian [Eq.~\eqref{eq:adiabaticHamiltonian}] for collisions of $^1$S-atoms with $^3\Sigma$-molecules can be found in Ref.~\cite{tscherbulDincao2023} except for the spin-spin interaction given by \cite{Krems:04}
\begin{equation}
    \hat{V}_{\text{SS}} = \frac{2}{3} \lambda_{\text{SS}} \sqrt{ \frac{ 4\pi }{ 5 } } \sqrt{6} \sum_q (-1)^q Y_{2 -q}(\hat{r}) [\hat{\bf S} \otimes \hat{\bf S}]_q^{(2)},
\end{equation}
where $\lambda_{\text{SS}}$ is the spin-spin interaction constant, $Y_{l m}(\hat{R})$ are spherical harmonics, and $[\hat{\bf S} \otimes \hat{\bf S}]_q^{(2)}$ is a tensor product.
The matrix elements in the TRAM basis are
\begin{align}
    \mel{(N L) J_r M_r}{ & \hat{V}_{\text{SS}} }{(N' L') J_r' M_r'} = \frac{2}{3} \lambda_{\text{SS}} \sqrt{ \frac{ 4\pi }{ 5 } } \sqrt{6} 
    \\ & \times \sum_q (-1)^q \mel{(N L) J_r M_r}{Y_{2 -q}(\hat{r})}{(N' L') J_r' M_r'} \mel{ S M_S' }{[\hat{\bf S} \otimes \hat{\bf S}]_q^{(2)} }{ S M_S' }.
    \label{eq:Vss_mel}
\end{align}
The matrix elements of the spherical harmonics in Eq.~\eqref{eq:Vss_mel} in the TRAM basis are \cite{tscherbulDincao2023}
\begin{equation}
    \mel{(N L) J_r M_r}{Y_{2 -q}}{(N' L') J_r' M_r'} = (-1)^{J_r - M_r} \threejm{Jr}{-M_r}{2}{-q}{J_r'}{M_r'} \mel{(N L) J_r}{|Y^{(2)}|}{(N' L') J_r'},
    \label{eq:mel_spherical_harmonics}
\end{equation}
where the reduced matrix element is given by
\begin{align}
    \mel{(N L) J_r}{|Y^{(2)}|}{(N' L') J_r'} = \delta_{L L'} (-1)^{L+J_r'} & \sqrt{ (2J_r + 1) (2J_r' + 1)} \sqrt{ (2N + 1) (2N' + 1)}
            \\ & \times \sixj{N}{J_r'}{J_r}{N'}{L}{2} \threejm{N}{0}{2}{0}{N'}{0}
    \label{eq:reducedmel_spherical_harmonics}
\end{align}
and the terms in the parenthesis and curly brackets are 3$j$ and 6$j$ symbols, respectively. 
The matrix elements of the tensor product of the electronic spin with itself are \cite{Krems:04}
\begin{equation}
    \mel{S M_S}{[\hat{\bf S} \otimes \hat{\bf S}]_q^{(2)}}{S M_S'} = (-1)^{S - M_S} \threejm{S}{-M_S}{2}{q}{S}{M_S'} \mel{S M_S}{|[\hat{\bf S} \otimes \hat{\bf S}]^{(2)}|}{S M_S'}
    \label{eq:mel_spin_tensor_product}
\end{equation}
The reduced matrix element is
\begin{equation}
    \mel{S}{| [\hat{\bf S} \otimes \hat{\bf S}]^{(2)} |}{S} = \sqrt{5} [(2S+1)S(S+1)] \sixj{1}{S}{1}{S}{2}{S}.
    \label{eq:reducedmel_spin_tensor_product}
\end{equation}
For $S = 1$, the 6$j$-symbol is equal to $1/6$, and the reduced matrix element is $\mel{S}{| [\hat{\bf S} \otimes \hat{\bf S}]^{(2)} |}{S} = \sqrt{5}$ \cite{Krems:04}.
Combining Eqs.~\eqref{eq:mel_spherical_harmonics}-\eqref{eq:reducedmel_spin_tensor_product} with Eq.~\eqref{eq:Vss_mel}, we arrive at the final expression for the matrix elements of the spin-spin interaction in the TRAM basis:
\begin{align}
    \mel{(N L) J_r M_r}{ & \hat{V}_{\text{SS}} }{(N' L') J_r' M_r'} = \delta_{L L'} (-1)^{J_r + J_r' - M_r + L + S - M_S} \frac{4 \sqrt{6\pi}}{3} \lambda_{\text{SS}} 
    \\ & \times \sqrt{ (2J_r + 1) (2J_r' + 1) (2N + 1) (2N' + 1)} \sixj{N}{J_r'}{J_r}{N'}{L}{2} \threejm{N}{0}{2}{0}{N'}{0}
    \\ & \times \sum_q (-1)^q \threejm{Jr}{-M_r}{2}{-q}{J_r'}{M_r} \threejm{S}{-M_S}{2}{q}{S}{M_S'}.
    \label{eq:final_Vss_mel}
\end{align}

\clearpage

\newpage

\appendix

\bibliography{cold_mol}

\end{document}